\documentclass[prl,showpacs,twocolumn,superscriptaddress]{revtex4-1}

\usepackage{amsfonts,amsmath,amssymb,graphicx}
\usepackage[usenames,dvipsnames]{xcolor}

\usepackage{pdfpages}

\newcommand{\be}{\begin{eqnarray}}
\newcommand{\ee}{\end{eqnarray}}

\newcommand{\bra}[1]{\langle{#1}|}
\newcommand{\ket}[1]{|{#1}\rangle}

\newcommand{\Ep}{\mathcal{E}_\textnormal{p}}
\newcommand{\E}{\mathcal{E}}

\newcommand{\hta}{\hat{a}}

\begin{document}
\title{Quantum annealing with a network of all-to-all connected, two-photon driven Kerr nonlinear oscillators}

\author{Shruti Puri}
\affiliation{Institut quantique and D\'{e}partment de Physique, Universit\'{e} de Sherbrooke, Sherbrooke, Qu\'{e}bec, Canada J1K 2R1}
\author{Christian Kraglund Andersen}
\affiliation{Department of Physics and Astronomy, Aarhus University, DK-8000 Aarhus, Denmark}
\author{Arne L. Grimsmo}
\affiliation{Institut quantique and D\'{e}partment de Physique, Universit\'{e} de Sherbrooke, Sherbrooke, Qu\'{e}bec, Canada J1K 2R1}
\author{Alexandre Blais}
\affiliation{Institut quantique and D\'{e}partment de Physique, Universit\'{e} de Sherbrooke, Sherbrooke, Qu\'{e}bec, Canada J1K 2R1}
\affiliation{Canadian Institute for Advanced Research, Toronto, Canada}

\begin{abstract}

Quantum annealing aims to solve combinatorial optimization problems mapped on to Ising interactions between quantum spins. A critical factor that limits the success of a quantum annealer is its sensitivity to noise, and intensive research is consequently focussed towards developing noise-resilient annealers. Here we propose a new paradigm for quantum annealing with a scalable network of all-to-all connected, two-photon driven Kerr-nonlinear resonators. Each of these resonators encode an Ising spin in a robust degenerate subspace formed by two coherent states of opposite phases. The fully-connected optimization problem is mapped onto local fields driving the resonators, which are themselves connected by local four-body interactions. We describe an adiabatic annealing protocol in this system and analyze its performance in the presence of photon loss. Numerical simulations indicate substantial resilience to this noise channel, making it a promising platform for implementing a large scale quantum Ising machine. Finally, we propose a realistic implementation of this scheme in circuit QED.

\end{abstract}
\maketitle
\section{Introduction}

Many hard combinatorial optimization problems arising in diverse areas such as physics, chemistry, biology, and social science~\cite{santoro2002theory,babbush2014adiabatic,perdomo2012finding,lucas2013ising} can be mapped onto finding the ground state of an Ising Hamiltonian. In general, finding the ground state of the Ising Hamiltonian, referred to as the Ising problem, is an NP-hard problem \cite{barahona1982computational}. Quantum annealing, based on adiabatic quantum computing (AQC)~\cite{farhi2000quantum,farhi2001quantum}, aims to find solutions to the Ising problem, with the hope of a significant speedup over classical algorithms. In AQC, a system is slowly evolved from the non-degenerate ground state of a trivial initial Hamiltonian to that of a final Hamiltonian encoding a computational problem. 
During the time-evolution, the energy spectrum of the system changes and for the adiabatic condition to be satisfied, the evolution must be slow compared to the inverse minimum energy gap between the instantaneous ground state and the excited states.
The scaling behavior of the gap with problem size, thus, determines the efficiency of the adiabatic annealing algorithm.

In order to perform quantum annealing, the Ising spins are mapped onto two levels of a quantum system, i.e. a qubit, and the optimization problem is encoded in the interactions between these qubits. Adiabatic optimization with a variety of physical systems such as nuclear magnetic spins~\cite{steffen2003experimental} and superconducting qubits~\cite{boixo2014evidence,barends2016digitized} has been demonstrated. However, despite great efforts, whether these systems are able to solve large problems in the presence of noise remains an open question~\cite{amin2009decoherence}. As a consequence, it is imperative to search for other physical systems with improved resilience to noise. A general Ising problem is defined on a fully connected graph of Ising spins, and efficient embedding of large problems with such long-range interactions in physical systems with local connectivity is a challenge. In one approach, a fully connected graph of Ising spins is embedded in a so-called Chimera graph~\cite{choi2008minor,choi2011minor}. 
Alternatively, a more recent embedding scheme was proposed by Lechner, Hauke and Zoller (LHZ)~\cite{lechner2015quantum} in which $N$ logical Ising spins are encoded in $M=N(N-1)/2$ physical spins with $M-N+1$ constraints. The physical spins represent the relative configuration of a pair of logical spins and an all-to-all connected Ising problem in the logical spins is realized by mapping the logical couplings onto local fields acting on the physical spins and a problem independent four-body coupling to enforce the constraints. While this scheme exhibits some intrinsic fault tolerance to weakly correlated errors~\cite{pastawski2016error}, better decoding strategies are required to enhance its performance in the presence of correlated spin-flip noise~\cite{albash2016simulated}. Nevertheless, the simple design requiring only precise control of local fields makes it attractive for scaling to large problem sizes.

In search of a physical platform for quantum annealing that is both scalable and has adequate robustness to noise, we propose to encode the Ising problem in a network of two-photon driven Kerr nonlinear resonators (KNRs). In our scheme, a single Ising spin is mapped onto two coherent states with opposite phases, which constitute a two-fold degenerate eigenspace of the two-photon-driven KNR in the rotating frame of the drive~\cite{puri2016engineering}. A similar encoding has been used to implement classical Ising machines in other systems~\cite{utsunomiya2011mapping,takata2012transient,wang2013coherent,marandi2014network}. In contrast, we propose to realize quantum adiabatic algorithms by encoding a quantum spin in the quasi-orthogonal coherent states. The dominant source of error in this system is single-photon loss from the resonators. However, since a coherent state is invariant under the action of the photon jump operator, the encoded Ising spin is stabilized against bit flips. We describe a circuit QED implementation of local magnetic fields and four-body coupling between such resonators to build a quantum annealing platform, where a fully connected graph of Ising spins is embedded using the LHZ scheme. The adiabatic optimization is carried out by initializing the resonators in the network to vacuum, and varying only single-site drives to evolve them to the correct ground state of the embedded Ising problem. Interestingly, as we demonstrate numerically with a single driven KNR, the probability for the system to jump from the instantaneous ground state to one of the excited states during the adiabatic protocol due to photon loss is greatly suppressed, when compared to conventional qubit implementations with equal noise strengths. This resilience to the detrimental effects of photon loss, combined with easy state initialization and final state detection by homodyne measurement of the resonators' field amplitudes, opens the door to realizing a large scale quantum annealer with favorable noise resistance.

\section{Results}
Quantum annealing is executed by evolving a system of $N$ spins under a time-dependent Hamiltonian
\be
\hat{H}(t)=\left(1-\frac{t}{\tau}\right)\hat{H}_i+\left(\frac{t}{\tau}\right)\hat{H}_p,
\label{AP}
\ee
where $\hat{H}_i$ is the initial trivial Hamiltonian whose ground state is known, and $\hat{H}_p$ is the final Hamiltonian at $t=\tau$ which encodes an Ising spin problem: 
$\hat{H}_p=~\sum_{i>j}^NJ_{i,j}\hat{\sigma}_{\mathrm{z},i}\hat{\sigma}_{\mathrm{z},j}$.
Here, $\hat{\sigma}_{\mathrm{z},i}=\ket{1}\bra{1}-\ket{0}\bra{0}$ is the Pauli-$z$ matrix for the $i^{\mathrm{th}}$ spin and $J_{i,j}$ is the interaction strength between the $i^\mathrm{th}$ and $j^\mathrm{th}$ spin. For simplicity, we have assumed a linear time dependence of the Hamiltonian in Eq.~\eqref{AP}, but more complex annealing schedules can be used. Crucially, the initial and final Hamiltonian do not commute $[\hat{H}_i,\hat{H}_p]\neq 0$. The system, initialized to the ground state of $\hat{H}_i$, adiabatically evolves to the ground state of the problem Hamiltonian, $\hat{H}_p$, at time $t=\tau$ if $\Delta_\mathrm{min}\tau\gg1$, where $\Delta_\mathrm{min}$ is the minimum energy gap~\cite{farhi2000quantum}.
In most existing schemes, the binary states of a qubit represent an Ising spin and the initial Hamiltonian is given by local transverse magnetic fields~\cite{boixo2016computational}. Here we show how quantum adiabatic annealing can be implemented by mapping the spin state $\{\ket{0},\ket{1}\}$ on two coherent states $\{\ket{-\alpha},\ket{\alpha}\}$ of a KNR. We, moreover, show that the time-dependent Hamiltonian can be implemented through simple single-site tunable drive fields.

\subsection{Single spin in a two-photon driven KNR}
\begin{figure*}
 \includegraphics[width=1.8\columnwidth]{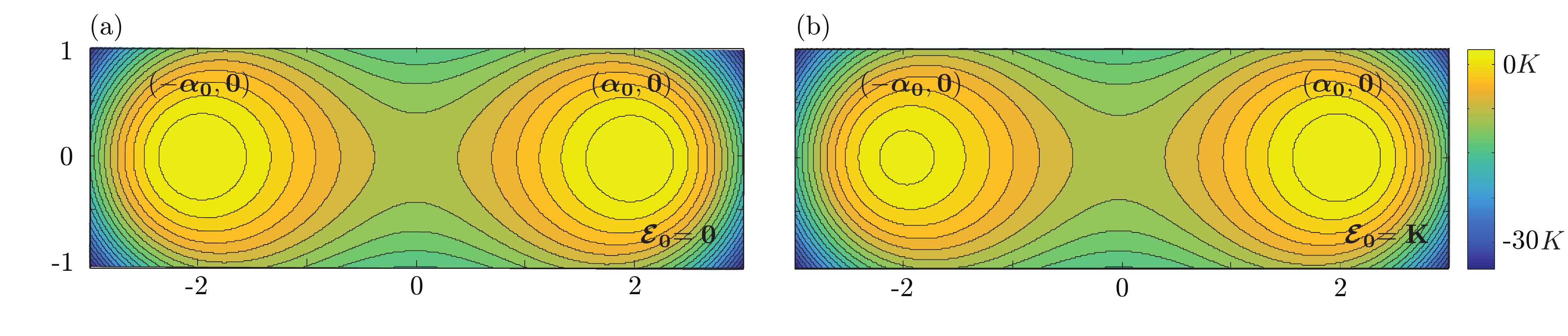}
 \caption{{\bf{Metapotential.}} Metapotential corresponding to $\hat{H}_p= -K\hta^{\dag 2}\hta^2+\Ep(\hta^{\dag 2}+\hta^2)+\E_0(\hta^\dag+\hta)$ with $\Ep=4K$ and (a) $\E_0=0$, (b) $\E_0=K$. These metapotentials are characterized by (a) two peaks of equal heights corresponding to the degenerate states $\ket{\bar{0}}$ and $\ket{\bar{1}}$, and (b) two peaks of different heights, indicating lifting of degeneracy between the encoded spin states $\ket{\bar{0}}$ and $\ket{\bar{1}}$.  \label{ClassPot}}
 \end{figure*}

The Hamiltonian of a two-photon driven KNR in the frame rotating at the frequency of the drive is given by, $\hat{H}_0=-K\hta^{\dag 2}\hta^2+\Ep(\hta^{\dag 2}+\hta^2)$, where $K$ is the Kerr-nonlinearity and $\Ep$ is the strength of the two-photon drive.
    The coherent states $\ket{\pm\alpha_0}$ are eigenstates of the photon annihilation operator $\hta\ket{\pm\alpha_0}=\pm\alpha_0\ket{\pm\alpha_0}$ and are stabilized in such a resonator with $\alpha_0=\sqrt{\Ep/K}$~\cite{puri2016engineering} (see also Methods). 
Intuitively, this is seen from the metapotential obtained by replacing the operators $\hta$ and $\hta^\dag$ with the complex classical variables $x+iy$ and $x-iy$ in the expression for $\hat{H}_0$~\cite{dykman2012fluctuating}. As shown in Fig.~\ref{ClassPot}(a), this metapotential features an inverted double well with two peaks of equal height at $(\pm\sqrt{\Ep/K},0)$. These are the two stable points in the metapotential (see Supplementary Note 1). This is consistent with the quantum picture according to which the coherent states $\ket{\pm\alpha_0}$ are two degenerate eigenstates of $\hat{H}_0$ with eigenenergy $\Ep^2/K$~\cite{puri2016engineering} (see also Methods). Having a well-defined two-state subspace, we choose to encode an Ising spin $\{\ket{\bar{0}},\ket{\bar{1}}\}$ in the stable states $\{\ket{-\alpha_0},\ket{\alpha_0}\}$. Importantly, if the rate of single photon loss ($\kappa$) is small $(\kappa\ll8\Ep)$, this mapping remains robust against single-photon loss from the resonator~\cite{puri2016engineering}. Moreover, the photon jump operator $\hta$ leaves the coherent states invariant $\hta\ket{\bar{0}/\bar{1}}=\pm \alpha_0\ket{\bar{0}/\bar{1}}$. As a result, if the amplitude $\alpha_0$ is large such that $\langle\bar{0}/\bar{1}|\hta|\bar{1}/\bar{0}\rangle=\mp\alpha_0e^{-2|\alpha_0|^2}\sim 0$, a single photon loss does not lead to a spin-flip error. 

Having defined the spin subspace, we now discuss the realization of a problem Hamiltonian in this system. As an illustrative example, we address the trivial problem of finding the ground state of a single spin in a magnetic field. Consider the Hamiltonian of a two-photon driven KNR with an additional weak single-photon drive of strength $\E_0$, $\hat{H}_p= -K\hta^{\dag 2}\hta^2+\Ep(\hta^{\dag 2}+\hta^2)+\E_0(\hta^\dag+\hta)$. The metapotential for this Hamiltonian, illustrated in Fig.~\ref{ClassPot}(b) for small $\E_0$, is an asymmetric inverted double well with peaks of unequal heights at $(\pm\alpha_0,0)$. Depending on if $\E_0>0$ or $\E_0<0$, the peak at $(-\alpha_0,0)$ is lower than the one at $(\alpha_0,0)$ or vice versa. These two states remain stable, but have different energies, indicating that the small single-photon field induces an effective magnetic field on the Ising spins $\{\ket{\bar{0}},\ket{\bar{1}}\}$. Indeed, a full quantum analysis shows that if $|\E_0|\ll4K|\alpha_0|^3$, then $\ket{\pm \alpha_0}$ remain the eigenstates of $\hat{H}_p$ but their degeneracy is lifted by $4\E_0\alpha_0$~\cite{puri2016engineering}. In other words, in the spin subspace, $\hat{H}_p$ can be expressed as $\hat{H}_p=2\E_0\alpha_0\hat{\bar{\sigma}}_\mathrm{z} + \mathrm{const.}$, with $\hat{\bar{\sigma}}_\mathrm{z}=\ket{\bar{1}}\bra{\bar{1}}-\ket{\bar{0}}\bra{\bar{0}}$, which is the required problem Hamiltonian
for a single spin in a magnetic field. Note that, if $\E_0$ increases then the eigenstates can deviate from coherent states (see Supplementary Note 2). Choosing $|\E_0|\ll4K|\alpha_0|^3$ ensures that $\{\ket{\bar{0}},\ket{\bar{1}}\}$ are indeed coherent states, so that $\langle\bar{0}/\bar{1}|\hta|\bar{1}/\bar{0}\rangle\sim 0$ and the encoded subspace is well protected from the photon loss channel. 

In correspondence with Eq.~\eqref{AP}, we require an initial Hamiltonian which does not commute with the final problem Hamiltonian and has a simple non-degenerate ground state. This is achieved by introducing a finite detuning $\delta_0>0$ between the drives and resonator. In a frame rotating at the frequency of the drives, the initial Hamiltonian is chosen as $\hat{H}_i=\delta_0\hta^\dag\hta-K\hta^{\dag 2}\hta^2$ with $\delta_0<K$. In this frame, the ground and first excited states are the vacuum $\ket{0}$ and single-photon Fock state $\ket{n=1}$ respectively, which are separated by an energy gap $\delta_0$. If a single-photon is lost from the resonator, the excited state $\ket{n=1}$ decays to the ground state $\ket{0}$ which, on the other hand, is invariant to photon loss. Since it is simple to prepare in the superconducting circuit realizations that we consider below, the vacuum state is a natural choice for the initial state.

\begin{figure*}
 \includegraphics[width=2\columnwidth]{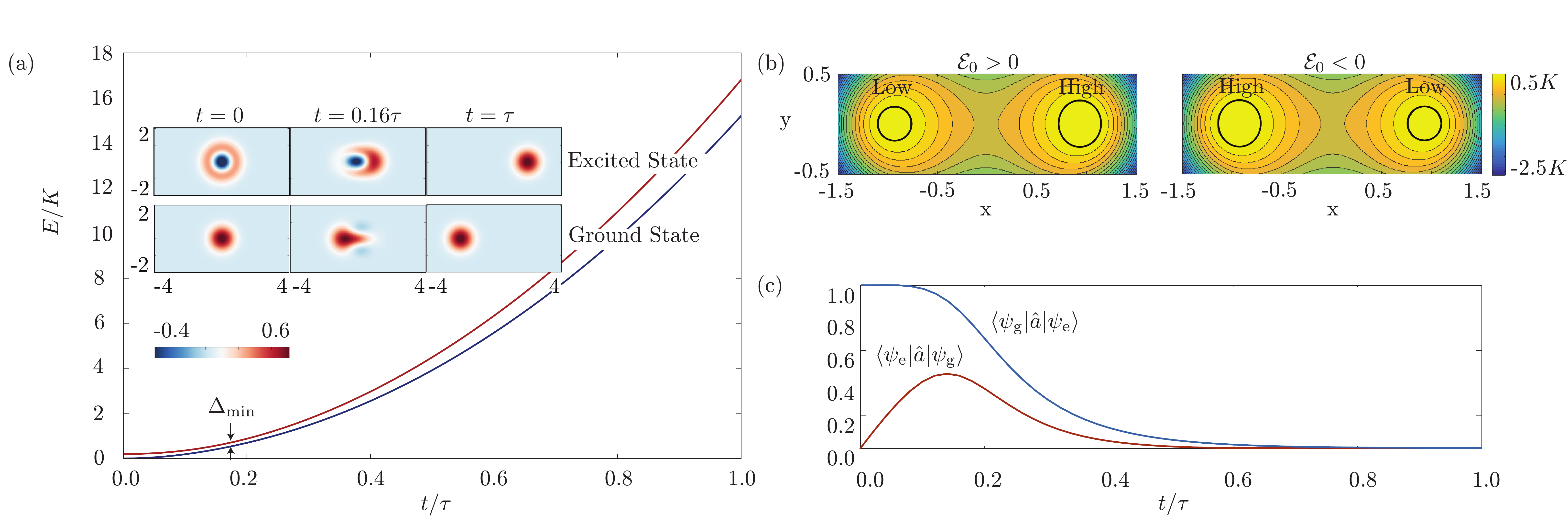}
 \caption{{\bf{Adiabatic protocol with single spin.}} Change of the energy of the ground and first excited state as a function of time in a single resonator for $\Ep=4K$, $\E_0=0.2K$ and $\delta_0=0.2K$. The minimum energy gap is also shown with $\Delta_\textrm{min}=0.16K$. The top and bottom panels in the inset show the Wigner function of the KNR state at three different times when initialized to either the excited $\ket{n=1}$ or (vacuum) ground state $\ket{0}$, respectively. (b) Metapotential corresponding to $\hat{H}_1(t=0.25\tau)$ with $\E_0=0.2K$ (left) and $E_0=-0.2K$ (right) showing two peaks of unequal height. The lower peak (corresponding to the ground state) is circular, whereas the higher one (corresponding to the excited state) is deformed as highlighted by black circles. (c) Transition matrix elements between the ground $\ket{\psi_g(t)}$ and excited states $\ket{\psi_e(t)}$ in the event of a photon jump during the adiabatic protocol. \label{eSingle} }
\end{figure*}

The time-dependent Hamiltonian required for the adiabatic computation can be realized by slowly varying the two- and single-photon drive strengths and detuning so that $\hat{H}_1(t)=(1-t/\tau)\hat{H}_i+(t/\tau)\hat{H}_p$ realizing Eq.~\eqref{AP} for a single-spin (see Methods). Note that the form of the Hamiltonian $\hat{H}_1(t)$ conveniently ensures that the nonlinear Kerr term is time-independent so that one only needs to vary the drives. By adiabatically controlling the frequency and amplitude of the drives it is possible to evolve the state of the KNR from the vacuum $\ket{0}$ at $t=0$, to the ground state of a single Ising spin in a magnetic field at $t=\tau$. Figure~\ref{eSingle}(a) shows the change of the energy landscape in time found by numerically diagonalizing the instantaneous Hamiltonian, $\hat{H}_1(t)$ for $\Ep=4K$, $\alpha_0={2}$, $\E_0=0.2K$ and $\delta_0=0.2K$. The minimum energy gap $\Delta_\mathrm{min}$ at the avoided level crossing is also indicated. As illustrated by the plots of the Wigner functions in the inset of Fig.~\ref{eSingle}(a), a resonator initialized to the vacuum state at $t=0$ evolves through highly non-classical and non-Gaussian states, towards the ground state $\ket{\bar{0}}$ at $t=\tau$, with $\tau\sim30/\Delta_\mathrm{min}$ in this example. If, on the other hand, the KNR is initialized to single-photon Fock state at $t=0$ then it evolves to the first excited state $\ket{\bar{1}}$ at $t=\tau$. The average probability to reach the correct ground state is $99.9\%$ for both $\E_0>0$ and $\E_0<0$. The $0.1\%$ probability of erroneously ending in the excited state arises from non-adiabatic errors and can be decreased by increasing the evolution time. For example, if $\tau=60/\Delta_\mathrm{min}$ then the success probability increases to 99.99$\%$. 
 
\subsection{Effect of single-photon loss}
An appealing feature of this implementation is that, at the start of the adiabatic protocol at $t=0$, the ground state (vacuum) is invariant under single-photon loss. Similarly, at the end of the adiabatic protocol at $t=\tau$, irrespective of the problem Hamiltonian (i.e., $\E_0>0$ or $\E_0<0$) the ground state (coherent states $\ket{\bar{0}}$ or $\ket{\bar{1}}$) is also invariant under single-photon loss. It follows that towards the beginning and end of the protocol, photon loss will not induce any errors. Moreover, we find that, even at intermediate times $0<t<\tau$, the ground state of $\hat{H}_1(t)$ remains largely unaffected by photon loss. This can be understood intuitively from the distortion of the metapotential, as shown in Fig.~\ref{eSingle}(b) for the example depicted in Fig.~\ref{eSingle}(a) at $t=0.25\tau$. The metapotential still shows two peaks, however, the region around the lower peak (corresponding to the ground state) is a circle whereas that around the higher peak (corresponding to the excited state) is deformed. This suggests that the ground state is closer to a coherent state and therefore more robust to photon loss than the excited state. Quantitatively, the effect of single-photon loss is seen by numerically evaluating~\cite{johansson2012qutip,johansson2013qutip} the transition matrix elements $\bra{\psi_g(t)}\hta\ket{\psi_e(t)}$, $\bra{\psi_e(t)}\hta\ket{\psi_g(t)}$ for the duration of the protocol, where $\ket{\psi_g(t)}$ and $\ket{\psi_e(t)}$ are the ground and excited state of $\hat{H}_1(t)$ respectively. As shown in Fig.~\ref{eSingle}(c) the transition from the ground to excited state is greatly suppressed throughout the whole adiabatic evolution. This asymmetry in the transition rates distinguishes the adiabatic protocol described here with two-photon driven KNRs from implementations with qubits~\cite{leib2016transmon}, something that will be made even clearer below with examples.


\subsection{Two coupled spins with driven KNRs}
Consider the problem of two interacting spins, which can be embedded in a system of two linearly coupled KNRs, each driven by a two-photon drive $\hat{H}_p=\sum_{k=1}^2\left[-K\hta^{\dag 2}_k\hta^2_k+\Ep(\hta^{\dag 2}_k+\hta^2_k)\right]+{J_{1,2}}(\hta^{\dag}_1\hta_2+\hta^\dag_2\hta_1)$. Here $J_{1,2}$ is the strength of the single-photon exchange coupling and, for simplicity, the two resonators are assumed to have identical parameters. For small ${J_{1,2}}$, this Hamiltonian can be expressed as 
$\hat{H}_p=4{J_{1,2}}|\alpha_0|^2\hat{\bar{\sigma}}_{\mathrm{z},1}\hat{\bar{\sigma}}_{\mathrm{z},2} + \text{const.}$, which is the required problem Hamiltonian~\cite{puri2016engineering}. The nature of the interaction, that is, ferromagnetic or anti-ferromagnetic, is encoded in the phase of the coupling ${J_{1,2}}<0$ or ${J_{1,2}}>0$, respectively.  For the initial Hamiltonian, we take $\hat{H}_i=\sum_{k}(\delta_0\hta^\dag_k\hta_k-K\hta^{\dag 2}_k\hta^2_k)+{J_{1,2}}(\hta^{\dag}_1\hta_2+\hta^\dag_2\hta_1)$. Following Eq.~\eqref{AP}, the full time-dependent Hamiltonian for the two-spin problem is $\hat{H}_2(t)=\left(1-t/\tau\right)\hat{H}_i+\left(t/\tau\right)\hat{H}_p$. 
Although it is possible to tune these parameters in time, with the above form of $\hat{H}_2(t)$, both the linear coupling and the Kerr nonlinearity are fixed during the adiabatic evolution.

\begin{figure}
 \includegraphics[width=\columnwidth]{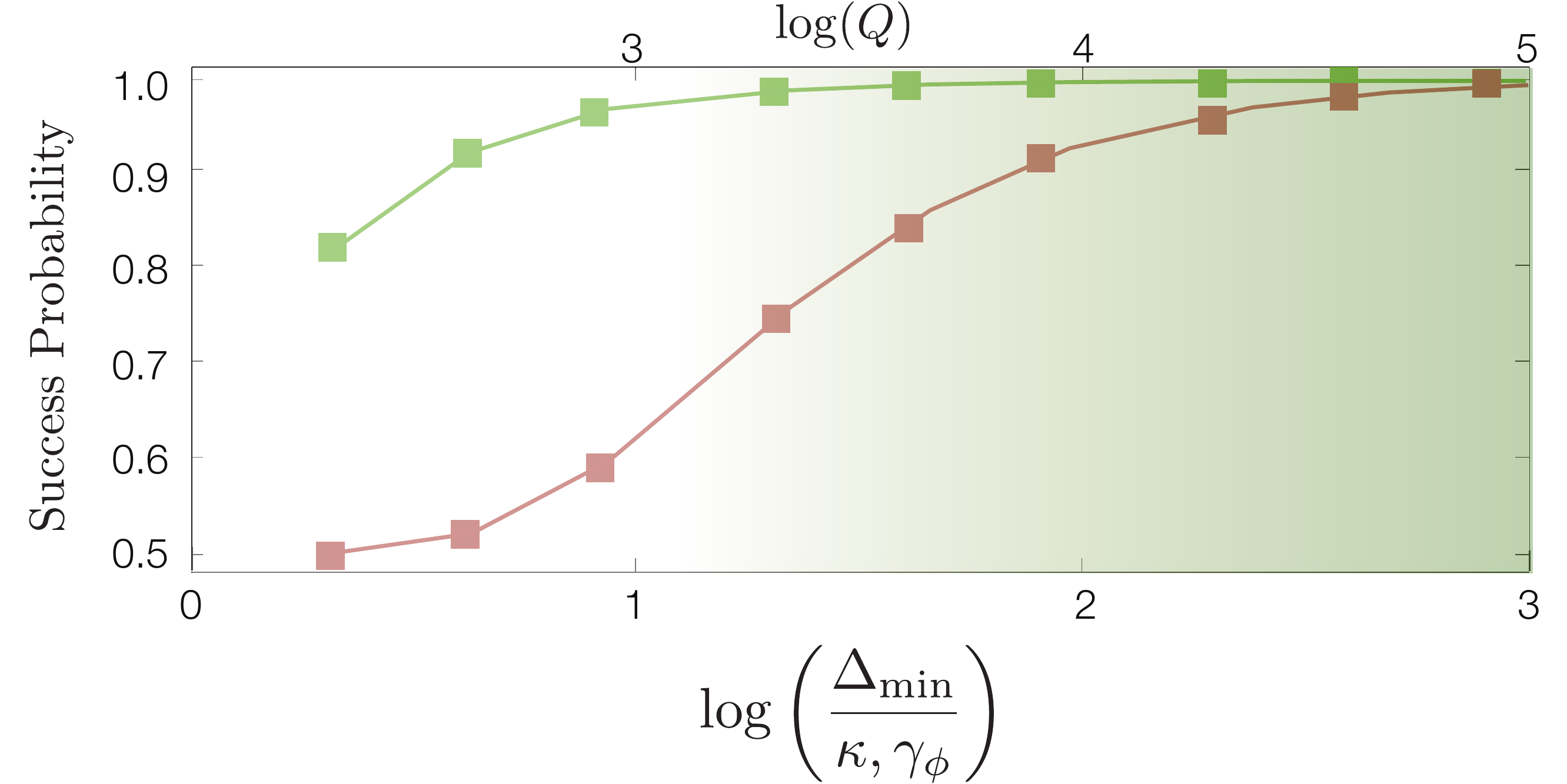}
 \caption{{\bf{Success probability for the two coupled spins problem.}} Loss-rate dependence of the success probability for the two-spin adiabatic algorithm in a system of two-photon driven KNRs with single-photon loss $\kappa$ (green squares) and qubits with pure dephasing at rate $\gamma_\phi$ (red squares). The quality factor $Q=\omega_\mathrm{r}/\kappa$ is indicated on the top axis for a KNR of frequency $\omega_\mathrm{r}/2\pi = 5$~GHz. \label{bar}}
 \end{figure}
 
Taking the initial detuning to be greater than the single-photon exchange rate, $\delta_0>{J_{1,2}}$, the ground state of $\hat{H}_2(0)$ is vacuum state. On the other hand, at $t=\tau$, $\{\ket{\bar{0},\bar{1}},\ket{\bar{1},\bar{0}}\}$ $(\{\ket{\bar{0},\bar{0}},\ket{\bar{1},\bar{1}}\})$ are the two degenerate ground states if the coupling is anti-ferromagnetic (ferromagnetic). Numerical simulations with both resonators initialized to vacuum shows that the coupled system reaches the entangled state $\mathcal{N}(\ket{\bar{0},\bar{1}}+\ket{\bar{1},\bar{0}})$ and $\mathcal{N}(\ket{\bar{0},\bar{0}}+\ket{\bar{1},\bar{1}})$, under anti-ferromagnetic and ferromagnetic coupling respectively. Here $\mathcal{N}=1/\sqrt{2(1+e^{-4|\alpha_0|^2})}$ is the normalization constant. For the parameters $\tau=50/\Delta_\mathrm{min}$, $\delta_0=K/4$, ${J_{1,2}}=K/10$ and $\Ep^0=2K$, so that $\alpha_0=\sqrt{2}$, the fidelity is $99.9\%$. Moreover, the probability that the system is in any one of the states $\ket{\bar{0}/\bar{1},\bar{0}/\bar{1}}$ is $99.99\%$, showing that the evolution is indeed restricted to this computational subspace. In the presence of single-photon loss, the coherence between the states is reduced. However the success probability (see Methods) to solve the Ising problem, which depends only on the diagonal elements of the density matrix (e.g. $\langle \bar{0},\bar{1}|\hat{\rho}(\tau)|\bar{0},\bar{1}\rangle$) remains high. For instance, with a large loss rate $\kappa=50/\tau$, the fidelity with respect to the superposition state $\mathcal{N}(\ket{\bar{0},\bar{0}}+\ket{\bar{1},\bar{1}})$ or $\mathcal{N}(\ket{\bar{0},\bar{1}}+\ket{\bar{1},\bar{0}})$ decreases to $37.6\%$, but the average success probability of the algorithm is $75.2\%$.

To characterize the effect of noise, a useful figure of merit is the ratio of the minimum energy gap to the loss rate ($\Delta_\mathrm{min}/\kappa$). The dependence of the average success probability on this ratio is presented in Fig.~\ref{bar} when the algorithm is implemented using KNRs (red squares) with single-photon loss ($\kappa$) or qubits (blue squares) with pure dephasing ($\gamma_\phi$). The success probability is averaged over all instances of the problem of two coupled spins (i.e., ferromagnetic and anti-ferromagnetic). All points are generated by varying the loss rates, while keeping $\Delta_\mathrm{min}$ and $\tau=20/\Delta_\mathrm{min}$ fixed. In the presence of pure dephasing, the success probability with qubits saturates to 50$\%$ for large $\gamma_\phi$. This is a consequence of the fact that the steady state of the qubits is an equal weight classical mixture of all possible computational states. On the other hand, with KNRs, in the presence of photon loss the rate at which the instantaneous ground state jumps to the excited state ($\propto\bra{\psi_e(t)}\hta\ket{\psi_g(t)}$) is small compared to the rate at which the instantaneous excited state jumps to the ground state ($\propto\bra{\psi_g(t)}\hta\ket{\psi_e(t)}$). For example, the success probability is $\sim75\%$ even when $\Delta_\mathrm{min}/\kappa\sim 1$. This shows that the algorithm implemented with two-photon driven KNRs has superior performance compared to that implemented with qubits in the presence of equal strength noise.

\begin{figure*}
\includegraphics[width=2\columnwidth]{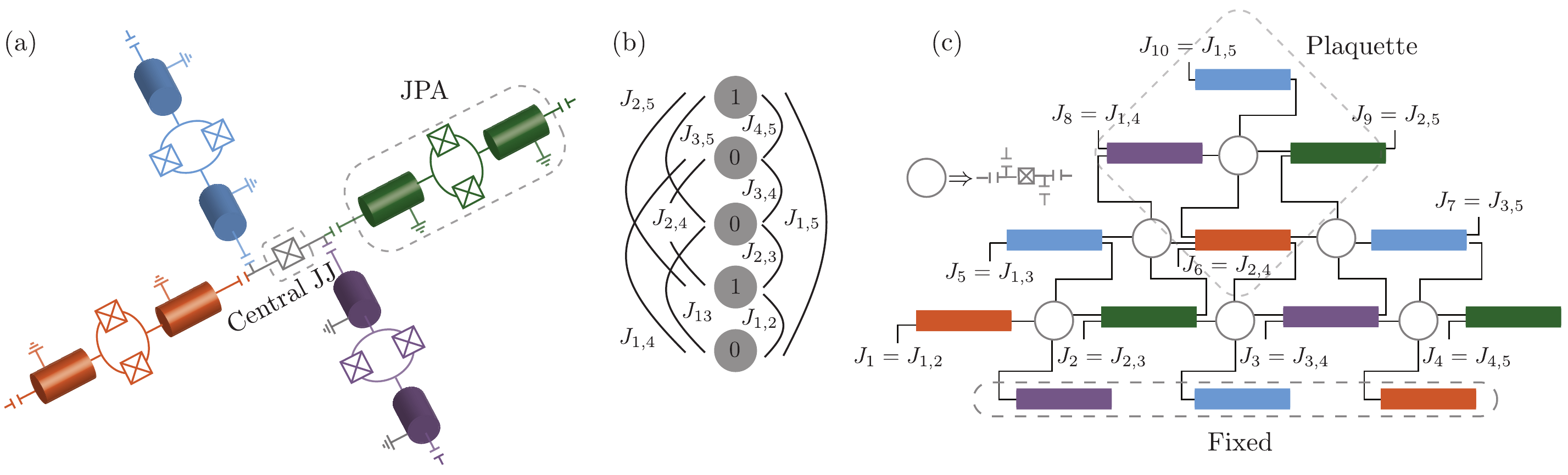}
\caption{{\bf{Physical realization of the LHZ scheme.}} (a) Illustration of the plaquette consisting of four JPAs coupled by a Josephson junction (JJ). The four JPAs have different frequencies (indicated by colors) and are driven by two-photon drives such that $\omega_{\mathrm{p},1}+\omega_{\mathrm{p},2}=\omega_{\mathrm{p},3}+\omega_{\mathrm{p},4}$. The nonlinearity of the JJ induces a four-body coupling between the KNRs. (b) Illustration of a fully-connected Ising problem with $N=5$ logical spins. (c) The same problem embedded on $M=10$ physical spins and $3$ fixed spins on the boundary.
\label{cartoon0}}
\end{figure*}

\subsection{All-to-all connected Ising problem with the LHZ scheme}
The above scheme can be scaled up with pairwise linear couplings in a network of KNRs, while still requiring only single-site drives. However, unlike the above one- and two-spin examples, most optimization problems of interest require controllable long-range interactions between a large number of Ising spins. 
Realizing such highly non-local Hamiltonian is a challenging hardware problem, but it may be solved by embedding schemes that map the Ising problem on a graph with only local interactions. As mentioned earlier, the LHZ scheme~\cite{lechner2015quantum} is one such technique in which the relative configuration of pairs of $N$ logical spins is mapped on $M=N(N-1)/2$ physical spins. A pair of logical spins in which both the spins are aligned $\ket{1,1}$ or $\ket{0,0}$ (or anti-aligned $\ket{1,0}$ or $\ket{0,1}$) is mapped on the two levels of the physical spin. The coupling between the logical pairs $J_{i,j}$ $(i=1,..N)$ is encoded in local magnetic fields on the physical spins $J_k$ $(k=1,..M)$. For a consistent mapping, $M-N+1$ energy penalties in the form of four-body coupling are introduced which enforce 
an even number of spin-flips around any closed loop in the logical spins. It was shown in Ref.~\cite{lechner2015quantum} that a fully connected graph can then be encoded in a planar architecture with only local connectivity.
The problem Hamiltonian in the physical spin basis becomes $\hat{H}_p^{\mathrm{LHZ},N}=\sum_{k=1}^M J_k\hat{\sigma}_{\mathrm{z},k}-\sum_{\langle i,j,k,l\rangle}C\hat{\sigma}_{\mathrm{z},i}\hat{\sigma}_{\mathrm{z},j}\hat{\sigma}_{\mathrm{z},k}\hat{\sigma}_{\mathrm{z},l}$ where $\langle i,j,k,l\rangle$ denotes the nearest-neighbor spins enforcing the constraint.

We now describe a circuit QED platform implementing the LHZ scheme by embedding the physical spins in the eigenbasis $\{\ket{\bar{0}},\ket{\bar{1}}\}$ of two-photon driven KNRs. Such a resonator can be realized as a microwave resonator terminated by a flux-pumped SQUID. The non-linear inductance of the SQUIDs induces a Kerr non-linearity, and a two-photon drive is introduced by flux-pumping at twice the resonator frequency. This is the exact same setup as is used for Josephson parametric amplifiers (JPAs), and we will therefore refer to this implementation of a Kerr nonlinear resonator as a JPA in the following~\cite{yamamoto:2008a,wustmann2013parametric,krantz2016single,puri2016engineering}. We envision the quantum annealing platform to be built with a group of four JPAs of frequencies $\omega_{\mathrm{r},i}$ ($i=1,2,3,4$) coupled to a single Josephson Junction (JJ) as shown in Fig.~\ref{cartoon0}(a). To obtain the time-dependent two-photon drive, the SQUID loop of each JPA is driven by a flux pump with tunable amplitude and frequency. The pump frequency is varied close to the resonator frequency, $\omega_{\mathrm{p},i}(t)\simeq2\omega_{r,k}$ (see Methods and Supplementary Note 4). Additional single-photon drives whose amplitude and frequency can be varied in time are also applied to each of the JPAs to provide the effective local magnetic field. Local four-body couplings are realized by the nonlinear inductance of the central JJ, see Supplementary Note 4. Choosing $\omega_{\mathrm{p},1}(t)+\omega_{\mathrm{p},2}(t)=\omega_{\mathrm{p},3}(t)+\omega_{\mathrm{p},4}(t)$ and taking the resonators to be detuned from each other, the central JJ induces a coupling of the form $-C(\hta^\dag_1\hta^\dag_2\hta_3\hta_4+h.c.)$ in the instantaneous rotating frame of the two-photon drives. This four-body interaction is an always-on coupling and its strength $C$ is determined by the JJ nonlinearity. Other circuits with tunable four-body coupling are possible, Supplementary Note 6. 
This group of four JPAs, which we will refer to as a plaquette, is the central building block of our architecture and can be scaled in the form of the triangular lattice required to implement the LHZ scheme. Lastly, the LHZ scheme also requires additional $N-2$ physical spins at the boundary that are fixed to the up state and which are implemented in our scheme as JPAs stabilized in the eigenstate $\ket{\bar{1}}$. As an illustration, Fig.~\ref{cartoon0}(b) depicts all the possible interactions in an Ising problem with $N=5$ logical spins and Fig.~\ref{cartoon0}(c) shows the corresponding triangular network of coupled KNRs. A final necessary component for a quantum annealing architecture is readout of the state of the physical spin. Here, this is realized by homodyne detection which can resolve the two coherent states $|\pm\alpha_0\rangle$ allowing the determination of the ground state configuration of the spins. 

In order to demonstrate the adiabatic algorithm for a non-trivial case, we embed on a plaquette a simple three-spin frustrated Ising problem, in which the spins are anti-ferromagnetically coupled to each other, $\hat{H}_p={J}\sum_{k,j=1,2,3}\hat{\sigma}_{\mathrm{z},k}\hat{\sigma}_{\mathrm{z},j}$ with $J>0$. This Hamiltonian has six degenerate ground states in the logical spin basis. Following the LHZ approach,
a mapping of $N=3$ logical spins requires $M=3$ physical spins (in our case 3 JPAs) and one physical spin fixed to up state (in our case a JPA initialized to the stable eigenstate $\ket{\bar{1}}$). Since, the physical spins $\{\ket{\bar{0}},\ket{\bar{1}}\}$ encoded in the JPAs constitute the relative alignment of the logical spins, there are three possible solutions in this basis $\ket{\bar{1},\bar{0},\bar{0}}$, $\ket{\bar{0},\bar{1},\bar{0}}$ and $\ket{\bar{0},\bar{0},\bar{1}}$. To implement the adiabatic protocol, the time-dependent Hamiltonian in a frame where each of the JPAs rotate at the instantaneous drive frequency is given by
\be \label{eq:plaq}
\begin{split}
\hat{H}^\mathrm{LHZ}(t)&=\left(1-\frac{t}{\tau}\right)\hat{H}_{i}+\left(\frac{t}{\tau}\right)\hat{H}^\mathrm{LHZ}_{p}+\hat{H}_\mathrm{fixed},
\end{split}
\ee
where
\be
\begin{split}
  \hat{H}_i&=\sum_{k=1}^3(\delta_0\hta^\dag_k\hta_k-K\hta^{\dag 2}_k\hta^2_k)-({C}\hta^\dag_1\hta^\dag_2\hta_3\hta_4+\mathrm{h.c.}),\\
  \hat{H}^\mathrm{LHZ}_p&=\sum_{k=1}^3\{-K\hta^{\dag 2}_k\hta^2_k+\Ep(\hta^{\dag 2}_k+\hta^2_k)+{J}(\hta^\dag_k+\hta_k)\}\\
  &-({C}\hta^\dag_1\hta^\dag_2\hta_3\hta_4+\mathrm{h.c.}),\\
\hat{H}_\mathrm{fixed}&=-K\hta^{\dag 2}_4\hta^2_4+\Ep(\hta^{\dag 2}_4+\hta^2_4).
\end{split}
\ee 
The anti-ferromagnetic coupling between the logical spins is represented by the single-photon drives on each JPA with amplitude ${J}>0$. At $t=0$, the ground state of this Hamiltonian is the vacuum $\ket{0,0,0}$. If the four-body coupling is weak then the problem Hamiltonian can be expressed as {$\hat{H}_p=2{J}\alpha_0\sum_{k=1}^3\hat{\bar{\sigma}}_\mathrm{z,i}-2{C}|\alpha_0|^4{\hat{\bar{\sigma}}_{\mathrm{z},1}}{\hat{\bar{\sigma}}_{\mathrm{z},2}}{\hat{\bar{\sigma}}_{\mathrm{z},3}}{\hat{\bar{\sigma}}_{\mathrm{z},4}} + \text{const.}$},
with $\alpha_0=\sqrt{\Ep/K}$. This realizes the required problem Hamiltonian in the LHZ scheme.

\begin{figure}
 \includegraphics[width=\columnwidth]{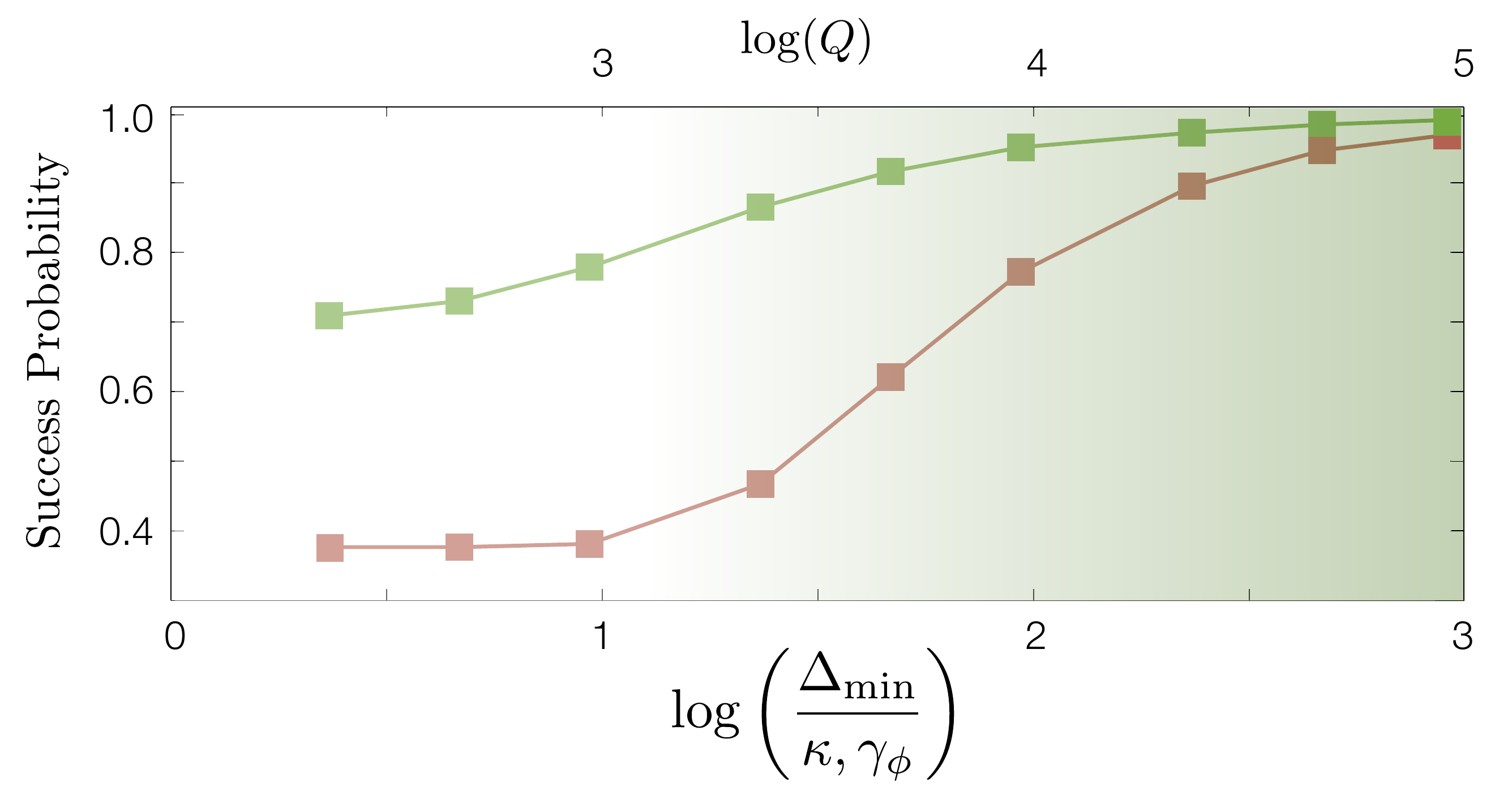}
 \caption{{\bf{Success probability for the frustrated three-spin problem.}} Probability of successfully finding the ground state of a frustrated three-spin Ising problem by implementing the adiabatic algorithm on a plaquette of four KNRs with single photon loss (green squares) for $\Ep^0=2K$, $\delta_0=0.45K$, $C=0.05K$, $J=0.095K$.  The success probability for an implementation with  qubits with pure dephasing rate $\gamma_\phi$ is also shown (red squares). The two cases are designed to have identical $\Delta_\mathrm{min}$ and computation time $\tau=40/\Delta_\mathrm{min}$. The quality factor $Q=\omega_\mathrm{r}/\kappa$ is indicated on the top axis for a KNR of frequency $\omega_\mathrm{r}/2\pi = 5$~GHz.
 \label{SP3spin}}
 \end{figure}

To illustrate the performance of this protocol, we numerically simulate the evolution subjected to the Hamiltonian in Eq.~\eqref{eq:plaq} with the three resonators initialized to vacuum and the fourth to the state $\ket{\bar{1}}$. With $\Ep=2K$, $\alpha_0=\sqrt{2}$, ${J}=0.095K$, ${C}=0.05K$, $\tau=40/\Delta_\mathrm{min}$ and $\kappa=0$, we find that the success probability to reach the ground state is $99.3\%$. The reduction in fidelity arises from the non-adiabatic errors. The probability for the system to be in one of the states $\ket{\bar{0}/\bar{1},\bar{0}/\bar{1},\bar{0}/\bar{1},\bar{0}/\bar{1}}$ is $99.98\%$ indicating that the final state is indeed restricted to  this subspace. Figure~\ref{SP3spin} shows the dependence of the success probability on single-photon loss rate (green). It also presents the success probability when the algorithm is implemented with qubits (red) subjected to dephasing noise (see Methods). Again, we find that the adiabatic protocol with JPAs (or two-photon driven KNRs) has superior performance in the presence of equal strength noise. 

This example of a simple frustrated three-spin problem demonstrates the performance of a single plaquette. Embedding of large Ising problems would require more plaquettes connected together as shown in Fig.~\ref{cartoon0}. Nevertheless, even in a larger lattice, each JPA is connected to only four other JPAs, making it likely that the final state remains restricted to the encoded subspace spanned by the states $\ket{\bar{0}},\ket{\bar{1}}$. Notably, the fully connected Ising problem is realized in the frame rotating at the drive frequencies, which is typically of the order of 5-10 GHz in experimental implementations. The thermal fluctuations around these frequencies are negligible in superconducting circuits operating at 10-30 mK. As a result thermal excitations in the reservoir are unlikely to drive the system out of the ground state even if the minimum energy gap is small. 

\section{Discussion}
We have introduced an adiabatic protocol performing quantum annealing with all- to-all connected Ising spins in a network of non-linear resonators with local interactions. The distinguishing feature of our scheme is that the spins are encoded in continuous-variable states of resonator fields. 
The restriction to two approximately orthogonal coherent states only happens in the late stage of the adiabatic evolution, and in general each site must be treated as a continuous variable system displaying rich physics, exemplified through non-Gaussian states, with negative-valued Wigner functions. How this behaviour persists in the presence of photon loss as the problem sizes are scaled up is an interesting question, as negativity of the Wigner function is directly related to classical non-simulability~\cite{PhysRevLett.109.230503,PhysRevLett.115.070501,PhysRevX.6.021039}.

An intriguing question is how the continuous variable nature of our system influences the annealer's computational capabilities when compared to a more conventional approach based on two-level systems evolving under a transverse field Ising Hamiltonian, i.e., where Eq.~\eqref{AP} is built from $H_i= \sum_i \hat{\sigma}_{x,i}$ and $H_p=\sum_{i,j} J_{i,j}\hat{\sigma}_{z,i}\hat{\sigma}_{z,j}$~\cite{boixo2014evidence,denchev2015computational}. For instance, we showed how the nature of the of quantum fluctuations around the instantaneous ground and excited states leads to increased stability of the ground state. As the size of the system increases, these continuous variable states might alter the nature of phase transitions during the adiabatic evolution, potentially leading to computational speedups~\cite{suzuki2007quantum,seki2015quantum}. It is also worth pointing out that our circuit QED implementation easily allows for adding correlated phase fluctuations given by interaction terms like $\hta^\dag_i\hta_i\hta^\dag_j\hta_j$ (see Supplementary Note 5). These terms do not affect the energy spectrum of the encoded problem Hamiltonian, but may modify the scaling of the minimal gap during the annealing protocol.

Another appealing feature that motivates further study into the complexity of our protocol is that the time-dependent Hamiltonian we use is generically non-stoquastic in the number basis. A stoquastic Hamiltonian by definition only has real, non-positive off-diagonal entries~\cite{bravyi2008complexity}, and the significance of this is that Hamiltonians in this class are directly amenable to quantum Monte Carlo simulations (stoquastic Hamiltonians do not have the so-called ``sign problem''). As an example, the transverse field Ising Hamiltonian is stoquastic. In contrast, our Hamiltonian has off-diagonal terms $\sum_k J_k(\hta_k^\dagger + \hta_k)$ in the LHZ embedding (or $\sum_{ij} J_{i,j}\hta_i\hta_j^\dagger + \text{h.c.}$ if this embedding is not used) with problem dependent signs (note that a simple diagonalization does not solve the problem due to the presence of quartic terms). The same is true if one considers matrix elements in the over-complete coherent state basis, $\langle \alpha| \hat{H}(t)|\beta \rangle$. It therefore does not appear straightforward to adapt quantum Monte Carlo techniques to this system.


Ultimately, further investigation into the performance of our adiabatic protocol on larger problem size is warranted. Currently, the large Hilbert space size prevents numerically exact simulations with more than a few resonators. Nonetheless, the results here strongly suggest that the adiabatic protocol with two-photon driven KNRs has excellent resistance to photon loss and thermal noise. Together with the highly non-classical physics displayed during the adiabatic evolution, this motivates the realization of a robust, scalable quantum Ising machine based on this architecture.

\section{Methods}
\subsection{Eigen-subspace of a two-photon driven KNR:}

Following Ref.~\cite{puri2016engineering}, the Hamiltonian of the two-photon driven KNR can be expressed as
\begin{align}
\begin{split}
\hat{H}_0&=-K\hta^{\dag 2}\hta^2+\Ep(\hta^{\dag 2}+\hta^2)\\
&=-K\left(\hta^{\dag 2}-\frac{\Ep}{K}\right)\left(\hta^2-\frac{\Ep}{K}\right)+\frac{\Ep^2}{K}\label{eq:HCassinian}.
\end{split}
\end{align}
This form form makes it clear that the two coherent states $\ket{\pm\alpha_0}$ with $\alpha_0=\sqrt{\Ep/K}$, which are the eigenstates of the annihilation operator $\hta$, are also degenerate eigenstates of Eq.~\eqref{eq:HCassinian} with energy $\Ep^2/K$. 

\subsection{Time-dependent Hamiltonian in the instantaneous rotating frame:}

We describe the required time-dependence of the amplitude and frequency of the drives to obtain the time-dependent Hamiltonians needed for the adiabatic evolution. As an illustration, consider the example of a two-photon driven KNR with additional single-photon drive whose Hamiltonian is written in the laboratory frame as
\begin{align}
\begin{split}
\hat{H}_\mathrm{1,Lab}(t)&=\omega_\mathrm{r}\hta^\dag\hta-K\hta^{\dag 2}\hta^2\\
&+\Ep(t)[e^{-i\omega_\mathrm{p}(t)t}\hta^{\dag 2}+e^{i\omega_\mathrm{p}(t)t}\hta^{2}]\\
&+\E_0(t)[e^{-i\omega_\mathrm{p}(t)t/2}\hta^\dag+e^{i\omega_\mathrm{p}(t)t/2}\hta].
\end{split}
\end{align}
Here, $\omega_\mathrm{r}$ is the fixed KNR frequency and $\omega_\mathrm{p}(t)$ is the time-dependent two-photon drive frequency. The frequency of the single-photon drive, of amplitude $\E_0(t)$, is chosen to be $\omega_\mathrm{p}(t)/2$ such that it is on resonance with the two-photon drive. In a rotating frame defined by the unitary transformation $\hat{U}=\exp[i\omega_\mathrm{p}(t)t~\hta^\dag\hta/2]$, this Hamiltonian reads
\begin{align}
\hat{H}_1(t)&=\hat{U}(t)^\dag\hat{H}_\mathrm{Lab}(t)\hat{U}(t)-i\dot{\hat{U}}(t)^\dag\hat{U}(t), \\
&=\left(\omega_{\mathrm{r}}-\frac{\omega_\mathrm{p}(t)}{2}-\dot{\omega}_\mathrm{p}(t)\frac{t}{2}\right)\hta^\dag\hta-K\hta^{\dag 2}\hta^2\\&+\Ep(t)(\hta^{\dag 2}+\hta^2)+\E_0(t)(\hta^\dag+\hta).
\end{align}
Choosing the time dependence of the frequency as $\omega_\mathrm{p}(t)=2\omega_\mathrm{r}-2\delta_0(1-t/2\tau)$, and drive strengths as $\Ep(t)=\Ep t/\tau$ and $\E_0(t)=\E_0 t/\tau$, the above Hamiltonian simplifies to
\begin{align}
\begin{split}
\hat{H}_1(t)&=\delta_0\left(1-\frac{t}{\tau}\right)\hta^\dag\hta-K\hta^{\dag 2}\hta^2+\left(\frac{t}{\tau}\right)\Ep(\hta^{\dag 2}+\hta^2)\\
&+\left(\frac{t}{\tau}\right)\E_0(\hta^\dag+\hta)\\
&=\left(1-\frac{t}{\tau}\right)(\delta_0\hta^\dag \hta-K\hta^{\dag 2}\hta^2)\\
&+\left(\frac{t}{\tau}\right)[-K\hta^{\dag 2}\hta^2+\Ep(\hta^{\dag 2}+\hta^2)+\E_0(\hta^\dag+\hta)].
\end{split}
\end{align}
This has the standard form of a linear interpolation between an initial Hamiltonian and a problem Hamiltonian that is required to implement the adiabatic protocol. 

As a second illustration, the time-dependent Hamiltonian for finding the ground state of a frustrated three spin problem embedded on a plaquette is (see Supplementary Note)
\begin{align}
\begin{split}
\hat{H}^\mathrm{LHZ}_\mathrm{Lab}(t)&=\sum_{k=1}^4(\omega_{\mathrm{r},k}\hta^\dag_k\hta_k-K\hta^{\dag 2}_k\hta^2_k)-C(\hta^\dag_1\hta^\dag_2\hta_3\hta_4+\mathrm{h.c.})\\
&+\sum_{k=1}^3J(t)[e^{-i\omega_{\mathrm{p},k}(t)t/2}\hta^{\dag}+e^{i\omega_{\mathrm{p},k}(t)t/2}\hta]\\
&+\sum_{k=1}^3\Ep(t)[e^{-i\omega_{\mathrm{p},k}(t)t}\hta^{\dag 2}+e^{i\omega_{\mathrm{p},k}(t)t}\hta^{2}]\\
&+\Ep[e^{-i\omega_{\mathrm{p},4}t}\hta^{\dag 2}+e^{i\omega_{\mathrm{p},4}t}\hta^{2}],
\end{split}
\end{align} 
where $\omega_{\mathrm{r},k}$ are the fixed resonator frequencies, $\omega_{\mathrm{p},k}(t)$ are the time-dependent two-photon drive frequencies. The resonators labelled $k=1$, 2 and 3 are driven by time-dependent two-photon and single-photon drives of strengths $\Ep(t)$, $J(t)$ and frequency $\omega_{\mathrm{p},k}(t)$, $\omega_{\mathrm{p},k}(t)/2$, respectively. On the other hand, the frequency and strength of the two-photon drive on the $k=4$ resonator is fixed. Applying the unitary $\hat{U}=\exp[i\sum_{k=1}^3\omega_{\mathrm{p},k}(t)t~\hta^\dag_k\hta_k/2]$ leads to the transformed Hamiltonian
\begin{align}
\begin{split}
\hat{H}^\mathrm{LHZ}(t)&=\sum_{k=1}^3\left(\omega_{\mathrm{r},k}-\frac{\omega_{\mathrm{p},k}(t)}{2}-\dot{\omega}_{\mathrm{p},k}(t)\frac{t}{2}\right)\hta^\dag_k\hta_k\\
&-K\hta^{\dag 2}_k\hta^2_k+J(t)(\hta^\dag_k+\hta_k)+\Ep(t)(\hta^{\dag 2}_k+\hta^{2}_k)\\
&-C(\hta^\dag_1\hta^\dag_2\hta_3\hta_4e^{i(\omega_{\mathrm{p},1}(t)+\omega_{\mathrm{p},2}(t)-\omega_{\mathrm{p},3}(t)-\omega_{\mathrm{p},4})t/2}\\
&+\mathrm{h.c.})\\
&+\left(\omega_{\mathrm{r},4}-\frac{\omega_{\mathrm{p},4}}{2}\right)\hta^\dag_4\hta_4-K\hta^{\dag 2}_4\hta^2_4+\Ep(\hta^{\dag 2}_4+\hta^{2}_4).
\end{split}
\end{align} 
To realize Eq.~\eqref{eq:plaq} implementing the adiabatic algorithm on this plaquette, we choose the drive frequencies such that $\omega_{\mathrm{p},k}(t)=2\omega_{\mathrm{r},k}-2\delta_0(1-t/2\tau)$ and $\omega_{\mathrm{p},4}=2\omega_{\mathrm{r},4}$ with their sum respecting $\omega_{\mathrm{p},1}(t)+\omega_{\mathrm{p},2}(t)=\omega_{\mathrm{p},3}(t)+\omega_{\mathrm{p},4}$. Moreover, we take the time-dependent amplitudes $\Ep(t)=\Ep t/\tau$ and $J(t)=J t/\tau$.

\subsection{Estimation of success probability:}

To estimate the success probability of the adiabatic algorithm with KNRs as shown by the green squares in Fig.~\ref{bar}, we numerically simulate the master equation $\dot{\hat{\rho}}=-[\hat{H}_2(t),\hat{\rho}]+\kappa\mathcal{D}[\hta_1]+\kappa\mathcal{D}[\hta_2]$, where the photon loss is accounted for by the Lindbladian $\mathcal{D}[\hta_i]=\hta_i\hat{\rho}\hta_i^\dag-(\hta^\dag_i\hta_i\hat{\rho}+\hat{\rho}\hta^\dag_i\hta_i)/2$~\cite{johansson2012qutip,johansson2013qutip}. 
It is important to keep in mind that even though the energy gap is small in the rotating frame, the KNRs laboratory frame frequencies $\omega_{\mathrm{r},k}$ are by for the largest energy scale. As a result, this standard quantum optics master equation correctly describes damping in this system~\cite{albash2015decoherence}. Moreover, because we are working with KNR frequencies in the GHz range, as is typical in supercondcuting circuits, thermal fluctuations are negligible. From his master equation, the success probability can be evaluated  as the probability of occupation of the correct ground state at the final time $t=\tau$, that is, $\bra{\bar{0},\bar{1}}\hat{\rho}(\tau)\ket{\bar{0},\bar{1}}+\bra{\bar{1},\bar{0}}\hat{\rho}(\tau)\ket{\bar{1},\bar{0}}$ and $\bra{\bar{0},\bar{0}}\hat{\rho}(\tau)\ket{\bar{0},\bar{0}}+\bra{\bar{1},\bar{1}}\hat{\rho}(\tau)\ket{\bar{1},\bar{1}}$ for $\E_0>0$ and $\E_0<0$, respectively. 

On the other hand, the master equation used to simulate the adiabatic algorithm with qubits is $\dot{\hat{\rho}}=-[\hat{H}^\mathrm{qubits}_2(t),\hat{\rho}]+\kappa\mathcal{D}[\hat{\sigma}_\mathrm{z,1}]+\kappa\mathcal{D}[\hat{\sigma}_\mathrm{z,2}]$, where
 \begin{align}
 \begin{split}
 \hat{H}^\mathrm{qubits}_2(t)&=\left(1-\frac{t}{\tau}\right) \hat{H}_i^\mathrm{qubits}+\left(\frac{t}{\tau}\right)\hat{H}_p^\mathrm{qubits}\\
 \hat{H}_i^\mathrm{qubits}&=U\sum_{i=1,2}\hat{\sigma}_\mathrm{x,i}, \quad \hat{H}_p^\mathrm{qubits}=J\hat{\sigma}_\mathrm{z,1}\hat{\sigma}_\mathrm{z,2},\\
 \mathcal{D}[\hat{\sigma}_\mathrm{z,i}]&=\gamma_\phi(\hat{\sigma}_\mathrm{z,i}\hat{\rho}\hat{\sigma}_\mathrm{z,i}-\hat{\rho}).
 \end{split}
 \end{align}
In this expression, $\hat{\sigma}_\mathrm{z,i}$ and $\hat{\sigma}_\mathrm{x,i}$ are Pauli operators in the computational basis formed by the ground $\ket{g}$ and excited state $\ket{e}$ of the $i^\mathrm{th}$ qubit. In these simulations, the qubits are initialized to the ground state of the initial transverse field, and the success probability (red squares in Fig.~\ref{bar}) is measured as the probability of occupation of the correct ground state at $t=\tau$, that is, $\bra{g,e}\hat{\rho}(\tau)\ket{g,e}+\bra{e,g}\hat{\rho}(\tau)\ket{e,g}$ and $\bra{g,g}\hat{\rho}(\tau)\ket{g,g}+\bra{e,e}\hat{\rho}(\tau)\ket{e,e}$ when $J>0$ and $J<0$, respectively. 

Finally, to obtain the data in Fig.~\ref{SP3spin} for the resonators (green squares), the simulated master equation is $\dot{\hat{\rho}}=-[\hat{H}^\mathrm{LHZ}(t),\hat{\rho}]+\sum_{i=1,2,3}\kappa\mathcal{D}[\hta_i]$ while, for qubits, it is $\dot{\hat{\rho}}=-[\hat{H}^\mathrm{LHZ,qubits}(t),\hat{\rho}]+\sum_{i=1,2,3}\gamma_\phi\mathcal{D}[\hat{\sigma}_\mathrm{z,i}]$. In these expressions,
 \begin{align}
 \begin{split}
 \hat{H}^\mathrm{LHZ,qubits}_2(t)&=\left(1-\frac{t}{\tau}\right) \hat{H}_i^\mathrm{qubits}+\left(\frac{t}{\tau}\right)\hat{H}_p^\mathrm{LHZ,qubits},\\
 \hat{H}_i^\mathrm{qubits}&=U\sum_{i=1,2,3}\hat{\sigma}_\mathrm{x,i}, \\
 \hat{H}_p^\mathrm{LHZ,qubits}&=J\sum\hat{\sigma}_\mathrm{z,i}+C\hat{\sigma}_\mathrm{z,1}\hat{\sigma}_\mathrm{z,2}\hat{\sigma}_\mathrm{z,3}\hat{\sigma}_\mathrm{z,4}.
 \end{split}
 \end{align}
The success probability is measured as the probability of occupation of the correct ground state at $t=\tau$, that is, $\bra{\bar{0},\bar{1},\bar{0}}\hat{\rho}(\tau)\ket{\bar{0},\bar{1},\bar{0}}+\bra{\bar{1},\bar{0},\bar{0}}\hat{\rho}(\tau)\ket{\bar{1},\bar{0},\bar{0}}+\bra{\bar{0},\bar{0},\bar{1}}\hat{\rho}(\tau)\ket{\bar{0},\bar{0},\bar{1}}$ (green squares in Fig.~\ref{SP3spin}) and $\bra{g,e,g}\hat{\rho}(\tau)\ket{g,e,g}+\bra{e,g,g}\hat{\rho}(\tau)\ket{e,g,g}+\bra{g,g,e}\hat{\rho}(\tau)\ket{g,g,e}$ (red squares in Fig.~\ref{SP3spin}).

 \bibliography{AQCpaper_v3.bbl}{}
 
 \clearpage


\section*{Supplementary material}

# SUPPLEMENTARY NOTE 1: STABILITY ANALYSIS OF THE META-POTENTIAL FOR A SINGLE TWO-PHOTON DRIVEN KNR

The coherent states $|\pm\alpha_0\rangle$, where $\alpha_0 = \sqrt{\mathcal{E}_\mathrm{p}/K}$, are eigenstates of the two-photon driven KNR Hamiltonian [1]

$$\hat{H}_0 = -K\hat{a}^{\dagger 2}\hat{a}^2 + \mathcal{E}_\mathrm{p}(\hat{a}^{\dagger 2} + \hat{a}^2). \tag{S1}$$

Following Ref. [1], we will outline a proof of this statement that will serve as the foundation for the analysis in the Supplementary Notes 2 and 3. We start by applying a displacement transformation $D(\alpha) = \exp(\alpha\hat{a}^\dagger - \alpha\hat{a})$ to $\hat{H}_0$, so that

$$\begin{aligned}\hat{H}_0' &= D^\dagger(\alpha)\hat{H}_0 D(\alpha) \\ &= [(-2K\alpha^2\alpha^* + 2\mathcal{E}_\mathrm{p}\alpha^*)\hat{a}^\dagger + \mathrm{h.c.}] + [(-K\alpha^2 + \mathcal{E}_\mathrm{p})\hat{a}^{\dagger 2} + \mathrm{h.c.}] - 4K|\alpha|^2\hat{a}^\dagger\hat{a} - K\hat{a}^{\dagger 2}\hat{a}^2 - (2K\alpha\hat{a}^{\dagger 2}\hat{a} + \mathrm{h.c.}).\end{aligned} \tag{S2, S3}$$

In the above expression, we have dropped the constant term $E(\alpha) = -K|\alpha|^4 + \mathcal{E}_\mathrm{p}^*\alpha^2 + \mathcal{E}_\mathrm{p}\alpha^{*2}$ which represents a shift in energy. We choose $\alpha$ such that the coefficient of $\hat{a}^\dagger$ satisfies $-2K\alpha^2\alpha^* + 2\mathcal{E}_\mathrm{p}\alpha^* = 0$, which is equivalent to finding the turning points of the metapotential of Fig. 1 of the manuscript. This equation has three solutions: $(0,0), (\pm\alpha_0, 0)$ corresponding to the dip and two peaks of the inverted double-well metapotential. At $(0,0)$, the Hamiltonian in Eq. (S3) represents a resonantly driven parametric amplifier. Therefore, in the absence of losses, large fluctuations make the system unstable around $(0,0)$ [2]. On the other hand, at $(\pm\alpha_0, 0)$, the Hamiltonian Eq. (S3) takes the form

$$\hat{H}_0'(\alpha = \pm\alpha_0) = -4K|\alpha_0|^2\hat{a}^\dagger\hat{a} - K\hat{a}^{\dagger 2}\hat{a}^2 - (\pm 2K\alpha_0\hat{a}^{\dagger 2}\hat{a} + \mathrm{h.c.}). \tag{S4}$$

With this normally ordered form, we immediately conclude that the vacuum $|0\rangle$ is an eigenstate of $\hat{H}_0'$ in the displaced frame. It immediately follows that the coherent states $|\pm\alpha_0\rangle$ are the eigenstates of the Hamiltonian $\hat{H}_0$ in the non-displaced frame. Furthermore, these are degenerate eigenstates as $E(\alpha_0) = E(-\alpha_0)$.

# SUPPLEMENTARY NOTE 2: THE EIGEN-SUBSPACE IN THE PRESENCE OF SINGLE-PHOTON DRIVE

The Hamiltonian of a two-photon driven KNR with an additional single-photon drive is given by

$$\hat{H} = -K\hat{a}^\dagger\hat{a}^\dagger\hat{a}\hat{a} + \mathcal{E}_\mathrm{p}(\hat{a}^{\dagger 2} + \hat{a}^2) + \mathcal{E}_0(\hat{a}^\dagger + \hat{a}). \tag{S5}$$

Under a displacement transformation $D(\alpha)$ this Hamiltonian reads

$$\begin{aligned}\hat{H}' &= [(-2K\alpha^2\alpha^* + 2\mathcal{E}_\mathrm{p}\alpha^* + \mathcal{E}_0)\hat{a}^\dagger + \mathrm{h.c.}] \\ &+ [(-K\alpha^2 + \mathcal{E}_\mathrm{p})\hat{a}^{\dagger 2} + \mathrm{h.c.}] - 4K|\alpha|^2\hat{a}^\dagger\hat{a} - K\hat{a}^{\dagger 2}\hat{a}^2 - (2K\alpha\hat{a}^{\dagger 2}\hat{a} + \mathrm{h.c.}),\end{aligned} \tag{S6}$$

where we have again dropped the constant term $E(\alpha) = -K|\alpha|^4 + \mathcal{E}_\mathrm{p}(\alpha^2 + \alpha^{*2}) + \mathcal{E}_0(\alpha + \alpha^*)$ representing a shift in energy. Following the same steps as above, the coefficient of the $\hat{a}^\dagger$ terms vanish if

$$-2K\alpha^2\alpha^* + 2\mathcal{E}_\mathrm{p}\alpha^* + \mathcal{E}_0 = 0. \tag{S7}$$

For small $\mathcal{E}_0$, this equation has three solutions of the form $(\pm\alpha_0 + \epsilon, 0), (\epsilon, 0)$ with $\epsilon = \mathcal{E}_0/4\mathcal{E}_\mathrm{p}$. In practice, we assume that the amplitude of the single-photon drive is small compared to that of the two-photon drive, $\mathcal{E}_0 \ll \mathcal{E}_\mathrm{p}$, so that $\epsilon \to 0$. If the condition Eq. (S7) is satisfied, then the Hamiltonian in the displaced frame reduces to

$$\hat{H}' = \left[-\frac{\mathcal{E}_0}{2\alpha^*}\hat{a}^{\dagger 2} + \mathrm{h.c.} - 4K|\alpha|^2\hat{a}^\dagger\hat{a}\right] - K\hat{a}^{\dagger 2}\hat{a}^2 - (2K\alpha\hat{a}^{\dagger 2}\hat{a} + \mathrm{h.c.}). \tag{S8}$$

Following Supplementary Note 1, $|0\rangle$ is an eigenstate of $\hat{H}'$ except for the first term which represents an off-resonant parametric drive of strength $|\mathcal{E}_0/2\alpha^*|$ detuned by $4K|\alpha|^2$. If $|\mathcal{E}_0/\alpha| \ll 4K|\alpha|^2$, fluctuations around $\alpha$ are small and $|0\rangle$ remains an eigenstate in the displaced frame. Of the three solutions to Eq. (S7), only $(\pm\alpha_0 + \epsilon, 0)$ satisfy the condition $|\mathcal{E}_0/(\pm\alpha_0 + \epsilon)| \ll 4K|\pm\alpha_0 + \epsilon|^2$. The third solution $(\epsilon, 0)$ is unstable because of the large quantum fluctuations around this point. As a result, in the non-displaced frame, the eigenstates of the system are $|\alpha_0 + \epsilon\rangle$ and $|-\alpha_0 + \epsilon\rangle$, where $\epsilon$ is a small correction. From the expression for $E(\alpha)$, it also clear that the degeneracy between the eigenstates $|\alpha_0 + \epsilon\rangle$ and $|-\alpha_0 + \epsilon\rangle$ is lifted by an amount $E(\alpha_0) - E(-\alpha_0) = 4\mathcal{E}_0\alpha_0$.

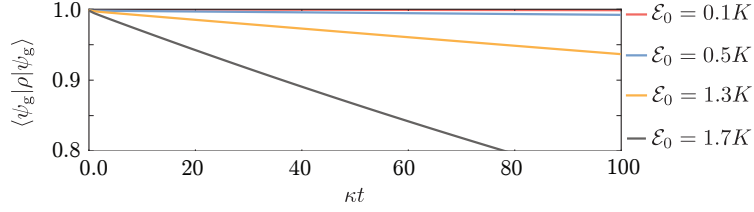

**Supplementary Figure 1. Ground state in presence of single-photon drive**: Time evolution of the probability of a single resonator to remain in the ground state $\langle\psi_g|\hat{\rho}|\psi_g\rangle$, for different single-photon drive strength. The two-photon drive strength is fixed to $\mathcal{E}_\mathrm{p} = 2K$ and $\kappa = 0.2K$.

Increasing $\mathcal{E}_0$, leads to squeezing due to the first term of Eq. (S8). As a result, the states are no longer coherent or, in other words, they are no longer the eigenstates of the photon annihilation operator $\hat{a}$ and hence are no longer protected against the single-photon loss channel. To illustrate this effect quantitatively, we numerically diagonalize the Hamiltonian in Eq. (S5) to evaluate the ground state for a fixed two-photon drive strength $\mathcal{E}_\mathrm{p} = 2K$ and variable strength single-photon drive amplitude. We numerically solve the master equation for the Hamiltonian in Eq. (S5), $\dot{\hat{\rho}} = -i[\hat{H}, \hat{\rho}] + \kappa[\hat{a}\rho\hat{a}^\dagger - (\hat{a}^\dagger\hat{a}\hat{\rho} + \hat{\rho}\hat{a}^\dagger\hat{a})/2]$ with the system initialized to the ground state at $t = 0$ and $\kappa = 0.2K$ [3, 4]. Finally, from these results, we compute the probability of the system to remain in the ground state $\langle\psi_g|\hat{\rho}|\psi_g\rangle$ after time $t$, presented in Supplementary Figure **1**. As seen from these numerical results, the ground state is invariant to single-photon loss for small $\mathcal{E}_0$, confirming our prediction from the simple theoretical analysis.

## SUPPLEMENTARY NOTE 3: EFFECT OF SINGLE PHOTON LOSS DURING THE ANNEALING PROTOCOL FOR A SINGLE DRIVEN KNR

In this section, we provide a simple theoretical analysis to understand the behaviour of the instantaneous ground and excited state during the annealing protocol with a single driven KNR in the presence of single-photon loss. The time-dependent Hamiltonian of this system is given by

$$\hat{H}_1(t) = \left(1 - \frac{t}{\tau}\right)\delta_0\hat{a}^\dagger\hat{a} - K\hat{a}^{\dagger 2}\hat{a}^2 + \left(\frac{t}{\tau}\right)[\mathcal{E}_\mathrm{p}(\hat{a}^{\dagger 2} + \hat{a}^2) + \mathcal{E}_0(\hat{a}^\dagger + \hat{a})]. \quad (S9)$$

In the presence of single-photon loss, the dynamics of the system is described by the master equation $\dot{\hat{\rho}} = -i[\hat{H}_{1,\mathrm{eff}}(t)\hat{\rho} - \hat{\rho}\hat{H}_{1,\mathrm{eff}}^\dagger(t)] + \kappa\hat{a}\rho\hat{a}^\dagger$, where $\kappa$ is rate of photon loss and $\hat{H}_{1,\mathrm{eff}}(t) = \hat{H}_1(t) - \kappa\hat{a}^\dagger\hat{a}/2$. Following Supplementary Notes 1 and 2, we apply a displacement transformation $D(\alpha)$ to $\hat{H}_{1,\mathrm{eff}}(t)$

$$\begin{aligned}\hat{H}'_{1,\mathrm{eff}} = &[(-2K\alpha^2\alpha^* + 2\mathcal{E}_\mathrm{p}(t)\alpha^* + \delta(t)\alpha - i\frac{\kappa}{2}\alpha + \mathcal{E}_0(t))\hat{a}^\dagger + \mathrm{h.c.}] \\ &+ [(-K\alpha^2 + \mathcal{E}_\mathrm{p})\hat{a}^{\dagger 2} + \mathrm{h.c.}] - 4K|\alpha|^2\hat{a}^\dagger\hat{a} + \delta(t)\hat{a}^\dagger\hat{a} - i\frac{\kappa}{2}\hat{a}^\dagger\hat{a} - K\hat{a}^{\dagger 2}\hat{a}^2 - (2K\alpha\hat{a}^{\dagger 2}\hat{a} + \mathrm{h.c.}).\end{aligned} \quad (S10)$$

For convenience, the variables have been redefined as $\delta(t) = \delta_0(1 - t/\tau)$, $\mathcal{E}_\mathrm{p}(t) = \mathcal{E}_\mathrm{p}t/\tau$ and $\mathcal{E}_0(t) = \mathcal{E}_0t/\tau$. Again, we take $\alpha$ to satisfy

$$-2K\alpha^2\alpha^* + 2\mathcal{E}_\mathrm{p}(t)\alpha^* + \delta(t)\alpha - i\frac{\kappa}{2}\alpha + \mathcal{E}_0(t) = 0, \quad (S11)$$

such that the effective Hamiltonian reads

$$\hat{H}'_{1,\mathrm{eff}} = [(-K\alpha^2 + \mathcal{E}_\mathrm{p})\hat{a}^{\dagger 2} + \mathrm{h.c.}] - 4K|\alpha|^2\hat{a}^\dagger\hat{a} + \delta(t)\hat{a}^\dagger\hat{a} - i\frac{\kappa}{2}\hat{a}^\dagger\hat{a} - K\hat{a}^{\dagger 2}\hat{a}^2 - (2K\alpha\hat{a}^{\dagger 2}\hat{a} + \mathrm{h.c.}). \quad (S12)$$

Eq. (S11) admits three solutions: $\sim (0, 0), (\pm\alpha'_0, 0)$, with $\alpha'_0 = \sqrt{[2\mathcal{E}_\mathrm{p}(t) + \delta(t)]/2K}$. Repeating the procedure described in Supplementary Note 1 and 2 we find, at short times when $\delta(t) \gg 2\mathcal{E}_\mathrm{p}(t)$, that $|0\rangle$ is approximately the lower energy, stable eigenstate of the Hamiltonian. As the evolution proceeds, the strength of the detuning and two-photon drive is

modified as $\delta(t) \ll 2\mathcal{E}_\mathrm{p}(t)$. In this case $(0,0)$ is not a stable eigenstate. On the other hand, at $(\pm\alpha'_0, 0)$ the Hamiltonian reads

$$\hat{H}'_{1,\mathrm{eff}} = \frac{1}{2}\left[\left\{-\left(\delta(t) - i\frac{\kappa}{2}\right)\frac{\alpha}{2\alpha'^*_0} - \frac{\mathcal{E}_0}{\alpha'^*_0}\right\}\hat{a}^{\dagger 2} + \mathrm{h.c.}\right] - 4K|\alpha'_0|^2\hat{a}^\dagger\hat{a} + \delta(t)\hat{a}^\dagger\hat{a} - i\frac{\kappa}{2}\hat{a}^\dagger\hat{a} - K\hat{a}^{\dagger 2}\hat{a}^2 - (2K\alpha'_0\hat{a}^{\dagger 2}\hat{a} + \mathrm{h.c.}). \tag{S13}$$

In the absence of the first term, $|0\rangle$ would be the eigenstate of the above Hamiltonian and hence the coherent states $|\pm\alpha'_0\rangle$ would be the eigenstates in the non-displaced frame. The first term represents a parametric drive whose amplitude has a magnitude $\zeta = |-(\delta(t) - i\kappa/2)(\alpha'_0/2\alpha'^*_0) - \mathcal{E}_0/\alpha'^*_0|$ and is detuned by $|-4K|\alpha'_0|^2\hat{a}^\dagger\hat{a} + \delta(t)\hat{a}^\dagger\hat{a} - i\kappa/2|$. In other words, the effect of detuning, photon-loss and single-photon drive is to squeeze the fluctuations around $(\pm\alpha'_0, 0)$. When $\delta(t) < 2\mathcal{E}_\mathrm{p}(t)$, $\kappa < 8\mathcal{E}_\mathrm{p}(t)$ and $\mathcal{E}(t) < 4K|\alpha'_0|^3$ then squeezing is negligible and the eigenstates in the non-displaced frame are approximately coherent states $|\pm\alpha'_0\rangle$. If $\mathcal{E}_0 > 0$, then $(-\alpha'_0, 0)$ corresponds to the lower energy state, and, on the other hand, if $\mathcal{E}_0 < 0$, $(\alpha'_0, 0)$ corresponds to the lower energy state. Note that the strength of squeezing for the lower energy state $|\zeta| = |-[\delta(t) - |\mathcal{E}(t)|/\sqrt{2\mathcal{E}_\mathrm{p}(t) + \delta(t)/2K}] - i\kappa/2|$ is smaller than that for higher energy state $|\zeta| = |-[\delta(t) + |\mathcal{E}(t)|/\sqrt{2\mathcal{E}_\mathrm{p}(t) + \delta(t)/2K}] - i\kappa/2|$ and hence, the deviation of lower energy state from the coherent state is smaller than the deviation of the higher energy state from the coherent state. This forms the basis for the observation that during the adiabatic evolution, the lower energy state which is the computational ground state is more stable against photon-jump operation.

## SUPPLEMENTARY NOTE 4: ENERGY SPECTRUM AND AVERAGE SUCCESS PROBABILITY OF PROBLEMS ENCODED ON A SINGLE PLAQUETTE

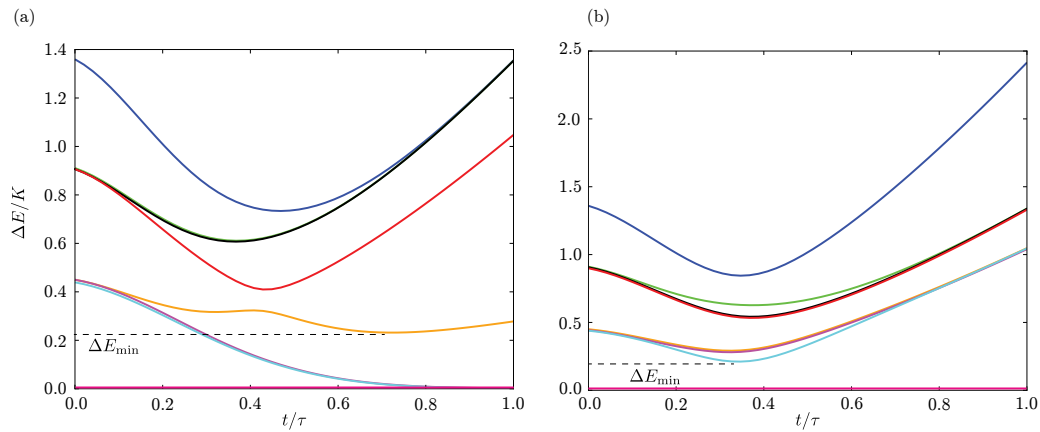

**Supplementary Figure 2. Evolution of the energy spectrum during the adiabatic evolution:** The figure shows the time-dependent energy spectrum during the adiabatic protocol for the fully connected three spin problem on a plaquette when the interactions between the spins is (a) anti-ferromagnetic, in which case the ground state of the problem in the physical spin basis is three fold-degenerate and (b) ferromagnetic, in which case the ground state of the problem in the physical spin basis is non-degenerate. The energy is measured with respect to the ground state $\Delta E = E_i - E_0$.

There are eight fully-connected Ising problems for three logical spins with weights $J_{ij} = \pm 1$: one where all spins are ferromagnetically coupled, one where all are anti-ferromagnetically coupled, three problems where two spins are ferromagnetically coupled, while the coupling to the third is anti-ferromagnetic and three problems where two spins are anti-ferromagnetically coupled, while the coupling to the third is ferromagnetic. All these problems can be encoded on a single plaquette by appropriately choosing the sign of the single photon drive. For example, as shown in the manuscript, the problem of all the logical spins coupled anti-ferromagnetically is embedded on the KNRs by applying single-photon drives such that $J > 0$. Similarly, to embed a problem where all logical spins are coupled ferromagnetically, all the applied single-photon drives must be such that $J < 0$. In order to embed the problem correctly in the LHZ scheme, the four-body coupling must be larger than local magnetic fields [5] and in our case, this implies $C|\alpha_0|^4 > J|\alpha_0|$. The adiabatic algorithm, as described in the manuscript, can be applied once the problem is correctly embedded.

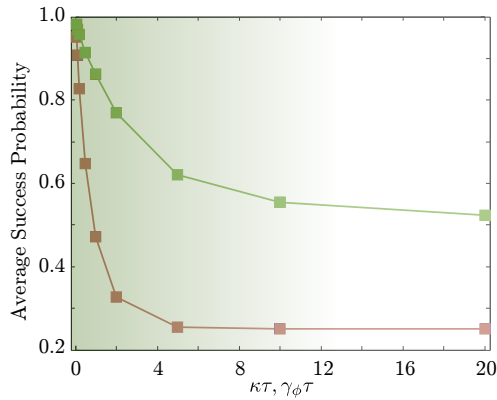

**Supplementary Figure 3**. **Average success probability for all problem instances on a plaquette:** The figure shows the dependence of the average success probability for all fully connected problems on a single plaquette when the adiabatic protocol is implemented using KNRs (green squares) characterized by single photon loss rate $\kappa$ and qubits (red squares) characterized by the dephasing rate $\gamma_\phi$. The total computation time $\tau$ is the same for both cases.

For example, when all the logical spins are coupled anti-ferromagnetically, the ground state is sixfold degenerate: $|0,0,1\rangle, |0,1,0\rangle, |1,0,0\rangle, |1,1,0\rangle, |1,0,1\rangle$ and $|0,1,1\rangle$. In the basis of physical spins, which map the relative configuration of the logical spins, the ground state is threefold degenerate $|\bar{0},\bar{0},\bar{1}\rangle, |\bar{0},\bar{1},\bar{0}\rangle$ and $|\bar{1},\bar{0},\bar{0}\rangle$. Supplementary Figure **2**(a) shows the evolution of the energy spectrum during the adiabatic protocol with $\mathcal{E}_\mathrm{p} = 2K$ and $C = 0.05K$. The single photon drive to all the resonators is $J = 0.095K$. The energy is measured from the ground state and the system starts from a non-degenerate ground state at $t = 0$. As expected, at $t = \tau$, the ground state becomes threefold degenerate.

If the logical spins are coupled ferromagnetically, then there are two possible ground states: $|0,0,0\rangle$ and $|1,1,1\rangle$, which means that in the physical spin basis there exists single non-degenerate ground state: $|\bar{1},\bar{1},\bar{1}\rangle$. Supplementary Figure **2**(b) shows the evolution of the energy spectrum during the adiabatic protocol with $\mathcal{E}_\mathrm{p} = 2K$ and $C = 0.05K$ but in this case the single photon drive is $J = -0.095K$ on all the resonators. The system starts from a non-degenerate ground state at $t = 0$ and indeed, at $t = \tau$, the ground state remains non-degenerate.

Similarly, it is possible to encode all eight problems on the plaquette and we find that in the absence of single-photon loss the average success probability to find the correct ground state with the adiabatic algorithm is 99.7% in time $\tau = 200/K$. Supplementary figure **3** shows the dependence of the success probability on the rate of single-photon loss. It also presents the success probability of the protocol implemented using qubits with dephasing rate $\gamma_\phi$. The time-dependent Hamiltonian for qubits is designed to have the same minimum energy gap as that with the JPAs and the duration of the protocol is also chosen to be the same. Clearly the adiabatic protocol with the JPAs outperform that with qubits in the presence of equal strength noise.

## SUPPLEMENTARY NOTE 4: PHYSICAL REALIZATION OF A PLAQUETTE

To implement the LHZ scheme [5] in our proposed setup, Josephson parametric amplifiers (JPAs) are arranged in a triangular lattice and coupled together a single coupling Josephson junction (JJ). As described in the main text, a single plaquette is comprised of four JPAs and the coupling JJ. The JPA is realized by embedding a SQUID in a resonator and modulating the flux through the SQUID at a frequency close to twice the frequency of the resonator [1, 6–8]. Recalling the Hamiltonian of the adiabatic protocol in the manuscript (Eq. (2) and Methods section in the manuscript), the flux modulation frequency is linearly varied from $\omega_{\mathrm{p},k}(0) = 2\omega_{\mathrm{r},k} - 2\delta_0$ at $t = 0$ to $\omega_{\mathrm{p},k}(\tau) = 2\omega_{\mathrm{r},k} - \delta_0$ at $t = \tau$. With an additional single-photon drive, the Hamiltonian of the $k^\mathrm{th}$ JPA can then be written as

$$\hat{H}_{\mathrm{JPA},k} = \omega_{\mathrm{r},k} \hat{a}_k^\dagger \hat{a}_k - K \hat{a}_k^{\dagger 2} \hat{a}_k^2 + J(t)[e^{-i\omega_{\mathrm{p},k}(t)t/2} \hat{a}^\dagger + e^{i\omega_{\mathrm{p},k}(t)t/2} \hat{a}] + \mathcal{E}_\mathrm{p}(t)[e^{-i\omega_{\mathrm{p},k}(t)t} \hat{a}^{\dagger 2} + e^{i\omega_{\mathrm{p},k}(t)t} \hat{a}^2]. \quad (\mathrm{S}14)$$

The Kerr nonlinearity $K$ and frequency $\omega_{\mathrm{r},k}$ are determined by the charging and Josephson energy of the junctions in the SQUID. By choosing appropriate parameters it is therefore possible to realize JPAs with different frequencies. The two-photon drive $\mathcal{E}_\mathrm{p}$ depends on the magnitude of the flux modulation. Consider Supplementary Figure **4** which shows four JPAs of different frequencies (indicated by different colors) capacitively coupled to a single JJ, which, as mentioned above, corresponds to a single plaquette and, thus, is the building block for the LHZ scheme. The

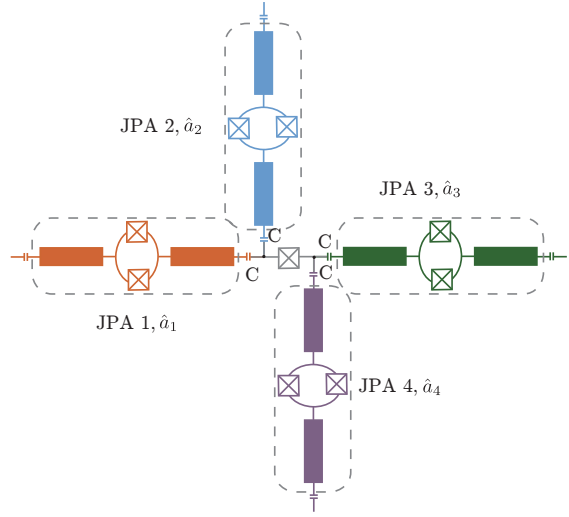

**Supplementary Figure 4. Illustration of the circuit for physical realization of the plaquette:** Four JPAs are linearly coupled to a mode of a Josephson junction via capacitors. The JPA modes are far detuned from the junction mode so that the coupling between them is dispersive. The non-linearity of the junction induces a four-body coupling between the JPAs if the frequencies of the two-photon drives to the JPAs are such that $\omega_{p,1} + \omega_{p,2} = \omega_{p,3} + \omega_{p,4}$.

Hamiltonian of this circuit is given by

$$\hat{H} = \sum_{k=1}^{4} \hat{H}_{\text{JPA},k} + \hat{H}_c, \tag{S15}$$

where the Hamiltonian of each JPA is given in Eq. (S14) and the coupling Hamiltonian is given by

$$\hat{H}_c = \omega_c \hat{a}_c^\dagger \hat{a}_c + g_1(\hat{a}_c^\dagger \hat{a}_1 + \hat{a}_1^\dagger \hat{a}_c) + g_2(\hat{a}_c^\dagger \hat{a}_2 + \hat{a}_2^\dagger \hat{a}_c) - g_3(\hat{a}_c^\dagger \hat{a}_3 + \hat{a}_3^\dagger \hat{a}_c) - g_4(\hat{a}_c^\dagger \hat{a}_4 + \hat{a}_4^\dagger \hat{a}_c)$$
$$- E_J\left(\cos\left(\frac{\hat{\Phi}}{\phi_0}\right) + \frac{1}{2}\left(\frac{\hat{\Phi}}{\phi_0}\right)^2\right), \quad \hat{\Phi} = \phi_c(\hat{a}_c^\dagger + \hat{a}_c), \tag{S16}$$

where $\phi_0 = \hbar/2e$ is the flux quantum, $\hat{a}_c$ ($\hat{a}_c^\dagger$) is the annihilation (creation) operator for the mode across the coupling JJ and $\phi_c$ is the standard deviation of the zero-point flux fluctuation for this JJ mode. $\hat{\Phi}$ is the phase across the junction, $E_J$ is its Josephson energy, $\omega_c$ is the frequency of the junction mode and $g_k$ is the rate at which energy is exchanged between this mode and the $k^{\text{th}}$ JPA. Following the definition for the mode operators, the quadratic phase term $\propto \hat{\Phi}^2$ has been removed from the cosine. In terms of the coupling capacitor $C$ and the zero point flux fluctuation for the JPA mode ($\phi_k$), the rate, $g_k = \phi_0^2 e^2/2C\phi_c\phi_k$. The total Hamiltonian is obtained by substituting Eq. (S16) and Eq. (S14) in Eq. (S15).

The frequency of the junction mode is designed to be largely detuned from the JPA modes, such that $\Delta_k = \omega_c - \omega_{r,k} \gg g_k$. It is then possible to apply a dispersive unitary transformation $\hat{U} = \exp(-i(g_1/\Delta_1)\hat{a}_c^\dagger \hat{a}_1 - i(g_2/\Delta_2)\hat{a}_c^\dagger \hat{a}_2 + i(g_3/\Delta_3)\hat{a}_c^\dagger \hat{a}_3 + i(g_4/\Delta_4)\hat{a}_c^\dagger \hat{a}_4 + \text{h.c.})$ to the total Hamiltonian, so that, to the second order in $g_k/\Delta_k$, the total Hamiltonian becomes

$$\hat{H} \sim \sum_{k=1}^{4}\left(\hat{H}_{\text{JPA},k} - \frac{g_k^2}{\Delta_k}\hat{a}_k^\dagger \hat{a}_k\right) + \left(\omega_c + \frac{g_k^2}{\Delta_k}\right)\hat{a}_c^\dagger \hat{a}_c - E_J\left(\cos\left(\frac{\hat{\Phi}'}{\phi_0}\right) + \frac{1}{2}\left(\frac{\hat{\Phi}'}{\phi_0}\right)^2\right), \tag{S17}$$

$$\sim \sum_{k=1}^{4}\left(\hat{H}_{\text{JPA},k} - \frac{g_k^2}{\Delta_k}\hat{a}_k^\dagger \hat{a}_k\right) + \left(\omega_c + \frac{g_k^2}{\Delta_k}\right)\hat{a}_c^\dagger \hat{a}_c - E_J \frac{1}{4!}\frac{\hat{\Phi}'^4}{\phi_0^4}, \tag{S18}$$

where

$$\hat{\Phi}' = \phi_c\left(\hat{a}_c^\dagger - \frac{g_1}{\Delta_1}\hat{a}_1^\dagger - \frac{g_2}{\Delta_2}\hat{a}_2^\dagger + \frac{g_3}{\Delta_3}\hat{a}_3^\dagger + \frac{g_4}{\Delta_4}\hat{a}_4^\dagger + \text{h.c.}\right). \tag{S19}$$

The expression in Eq. (S18) is obtained by expanding the cosine term to fourth order. The central Josephson junction mode is far detuned from the JPAs and is not driven externally so that we have $\langle \hat{a}_c \rangle = \langle \hat{a}_c^\dagger \hat{a}_c \rangle = 0$. Therefore it is possible to eliminate this mode and obtain a Hamiltonian only for the JPA modes. Furthermore, in a frame rotating at the frequencies of the two-photon drives, the photon annihilation operators transform as $\hat{a}_k \to e^{-i\omega_{p,k}(t)t}\hat{a}_k$. The two-photon drive frequencies are such that $\omega_{p,k} \neq \omega_{p,m}$ and $\omega_{p,1} + \omega_{p,2} = \omega_{p,3} + \omega_{p,4}$. As a result, we can eliminate fast rotating terms in the expansion of the last term in Eq. (S18) and realize the plaquette Hamiltonian

$$\hat{H}_{\text{plaquette}} \sim \sum_{k=1}^{4}\left(\hat{H}_{\text{JPA},k} - \frac{g_k^2}{\Delta_k}\hat{a}_k^\dagger \hat{a}_k\right) - E_J \frac{\phi_c^4}{\phi_0^4}\frac{g_1 g_2 g_3 g_4}{\Delta_1 \Delta_2 \Delta_3 \Delta_4}(\hat{a}_1^\dagger \hat{a}_2^\dagger \hat{a}_3 \hat{a}_4 + \text{h.c.}) - E_J \frac{\phi_c^4}{\phi_0^4}\sum_{k\neq m=1}^{4}\frac{g_k^2 g_m^2}{\Delta_k^2 \Delta_m^2}\hat{a}_k^\dagger \hat{a}_k \hat{a}_m^\dagger \hat{a}_m. \tag{S20}$$

The second part of the first term results in a frequency shift of the JPA modes due to off-resonant coupling with the JJ and only leads to a renormalization of the energies. The second term in the above expression is the desired four-body coupling between the JPAs. The four-body coupling strength defined in the manuscript can be written in terms of circuit parameters as $C = E_J \frac{\phi_c^4}{\phi_0^4}\frac{g_1 g_2 g_3 g_4}{\Delta_1 \Delta_2 \Delta_3 \Delta_4}$. As an example, choosing $E_J = 600/2\pi$ GHz, $\phi_c = 0.12\phi_0$, $g_k/\Delta_k \sim 0.12$ we estimate $C/2\pi = 63$ KHz, and for a typical strength of Kerr nonlinearity $K/2\pi = 600$ KHz, $C/K \sim 0.1$. The last term gives rise to a cross-Kerr interaction between the JPAs. As discussed in the following section, if the amplitude of the coherent states forming the computational subspace is large, then this term does not effect the structure of the energy spectrum.

### SUPPLEMENTARY NOTE 5: EFFECT OF THE CROSS-KERR COUPLING BETWEEN THE JPAS

In the computational subspace, the cross-Kerr coupling between the JPAs can be written as, $\hat{a}_k^\dagger \hat{a}_k \hat{a}_m^\dagger \hat{a}_m = |\alpha_0|^4 e^{-4|\alpha_0|^2}\hat{\sigma}_{k,x} \otimes \hat{\sigma}_{m,x} + \text{const.}$, where $\sigma_{k,x}$ is the Pauli operator $|\bar{1}\rangle\langle\bar{0}| + |\bar{0}\rangle\langle\bar{1}|$. If $\alpha_0$ is large then $e^{-4|\alpha_0|^2} \to 0$ and the cross-Kerr term only leads to a constant shift in energy without inducing any errors in the encoding. This can be confirmed by numerically evaluating the energy spectrum, for example, for the frustrated three spin problem on a single plaquette. If the problem is encoded correctly, we expect that at the end of the adiabatic protocol at $t = \tau$ the ground state is triple-fold degenerate $|\bar{0},\bar{0},\bar{1}\rangle, |\bar{0},\bar{1},\bar{0}\rangle, |\bar{1},\bar{0},\bar{0}\rangle$, where $|\bar{0}/\bar{1}\rangle = |\pm\alpha_0\rangle$. However, in presence of the cross-Kerr terms, we find that that the degeneracy is lifted by $\sim 0.005K$ if $\alpha_0 = \sqrt{2}$. If on the other hand, $\alpha_0 = \sqrt{3}$, then the lift in the degeneracy reduces to $0.0004K$. This confirms our prediction that as $\alpha_0$ increases, the errors due to the cross-Kerr terms decrease.

Furthermore, to ensure that the cross-Kerr terms do not induce additional dephasing in the presence of single-photon loss, we repeat the numerical simulations for the problem of three coupled spins embedded on a plaquette with these terms included and determine the average success probability. The numerical simulations are limited by the size of the Hilbert space and currently we simulate the adiabatic protocol with $\alpha_0 = \sqrt{2}$ by employing the following time-dependent Hamiltonian,

$$\hat{H} = \left(1 - \frac{t}{\tau}\right)\hat{H}_i + \left(\frac{t}{\tau}\right)\hat{H}_p \tag{S21}$$

where,

$$\hat{H}_i = \sum_{k=1}^{3}(\delta_0 \hat{a}_k^\dagger \hat{a}_k - K\hat{a}_k^{\dagger 2}\hat{a}_k^2) - (C\hat{a}_1^\dagger \hat{a}_2^\dagger \hat{a}_3 \hat{a}_4 + \text{h.c.}) - C\sum_{k,m=1}^{3}\hat{a}_k^\dagger \hat{a}_k \hat{a}_m^\dagger \hat{a}_m,$$

$$\hat{H}_p^{\text{LHZ}} = \sum_{k=1}^{3}\{(1-f)\delta_0 \hat{a}_k^\dagger \hat{a}_k - K\hat{a}_k^{\dagger 2}\hat{a}_k^2 + \mathcal{E}_p(\hat{a}_k^{\dagger 2} + \hat{a}_k^2) + J(\hat{a}_k^\dagger + \hat{a}_k)\} - (C\hat{a}_1^\dagger \hat{a}_2^\dagger \hat{a}_3 \hat{a}_4 + \text{h.c.}) - C\sum_{k,m=1}^{3}\hat{a}_k^\dagger \hat{a}_k \hat{a}_m^\dagger \hat{a}_m, \tag{S22}$$

$$\hat{H}_{\text{fixed}} = -K\hat{a}_4^{\dagger 2}\hat{a}_4^2 + \mathcal{E}_p(\hat{a}_4^{\dagger 2} + \hat{a}_4^2).$$

In contrast to Eq. (2) in the main text, we have included the cross-Kerr terms. In addition, we have included extra detuning to the problem Hamiltonian $\propto (1-f)$ in order to ensure a consistent encoding of the problem even in the presence of small $\alpha_0$. In the example presented here we use $f = 0.7$. The average success probability for all the eight possible problems on the plaquette, shown in Fig. **5**, effectively does not change compared to the case when the $\sum_{k,m=1}^{3}\hat{a}_k^\dagger \hat{a}_k \hat{a}_m^\dagger \hat{a}_m$ term is neglected.

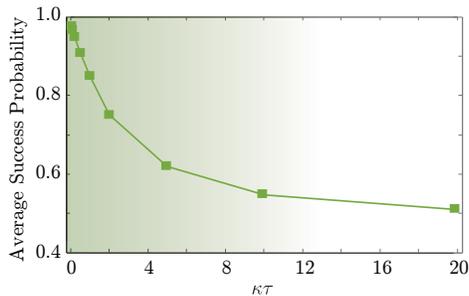

**Supplementary Figure 5**. **Average success probability for all problem instances on a plaquette with residual interactions:** The figure shows the dependence of the average success probability for all fully connected problems on a single plaquette with the $\sum_{k,m=1}^{3} \hat{a}_k^\dagger \hat{a}_k \hat{a}_m^\dagger \hat{a}_m$ term included.

## SUPPLEMENTARY NOTE 6: TUNABLE FOUR-BODY COUPLING WITH A JOSEPHSON RING MODULATOR

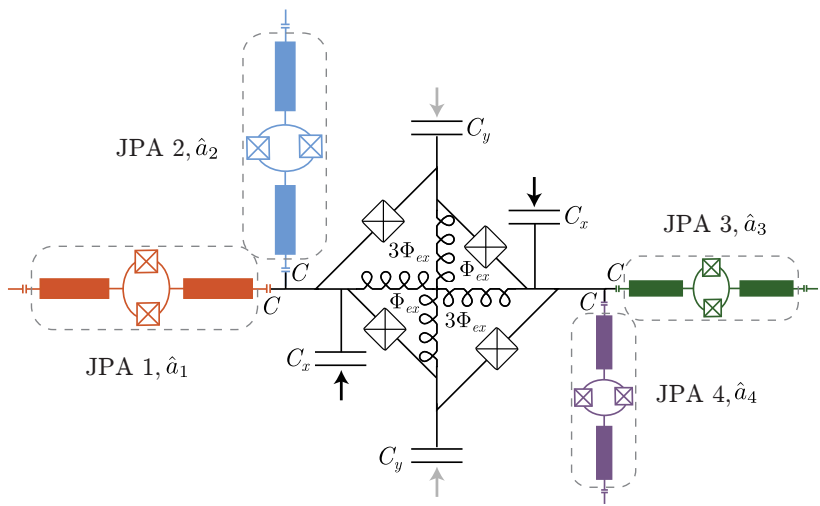

**Supplementary Figure 6**. **Tunable four-body coupling with JRM:** The figure illustrates the circuit for implementing a tunable four-body interaction using a Josephson ring modulator (JRM). Classical microwave drives of equal strength but opposite phase, as shown with the dark and light arrows, activates the four-body coupling between the JPAs.

An alternative way to implement a four-body interaction in a tunable way is by using an imbalanced shunted Josephson Ring Modulator (JRM) as illustrated in Supplementary Figure **6**. The JRM consists of 3 orthogonal mutually interacting modes $\hat{\phi}_x, \hat{\phi}_y$ and $\hat{\phi}_z$ [9, 10], which also couple with the JPA modes. The Hamiltonian can be expressed as

$$\hat{H}_c = \omega_x \hat{a}_x^\dagger \hat{a}_x + \omega_y \hat{a}_y^\dagger \hat{a}_y + \omega_z \hat{a}_z^\dagger \hat{a}_z + \sum_{\alpha=x,y,z} \sum_{i=1}^{4} g_i^\alpha (\hat{a}_i^\dagger \hat{a}_\alpha + \hat{a}_\alpha^\dagger \hat{a}_i) + U_J(\hat{\phi}_x, \hat{\phi}_y, \hat{\phi}_z). \tag{S23}$$

In the above expression, $\hat{a}_\alpha$ are the JRM mode operators and in terms of the zero point flux fluctuation, $\hat{\phi}_\alpha = \phi_\alpha(\hat{a}_\alpha + \hat{a}_\alpha^\dagger)$. The coupling rate, $g_i^\alpha$, is given by $g_i^\alpha = e^2 \phi_0^2/(2C\phi_\alpha\phi_i)$. The potential $U_J$ represents the interaction between the JRM

modes and is expressed as

$$U_J = -E_J \left[ 2\cos\left(\frac{\phi_x - \phi_y}{2\phi_0}\right) \cos\left(\frac{\phi_z}{\phi_0} + 3\Phi_{ex}\right) + 2\cos\left(\frac{\phi_x + \phi_y}{2\phi}\right) \cos\left(-\frac{\phi_z}{\phi_0} + \Phi_{ex}\right) \right], \tag{S24}$$

where $3\Phi_{ex}$ and $\Phi_{ext}$ are the external flux applied to the big and small loops respectively of the JRM. Note that we ought to subtract the second order term of each mode from the potential, but for convenience we omit these terms here as the correction to the frequencies are small. We now choose the external flux to be $\Phi_{ex} = \pi/2$, so that we have

$$U_J = -4E_J \cos\left(\frac{\phi_x}{2\phi_0}\right) \cos\left(\frac{\phi_y}{2\phi_0}\right) \sin\left(\frac{\phi_z}{\phi_0}\right). \tag{S25}$$

The above expression for $U_J$ is substituted in Eq. (S23) to obtain a complete expression for the coupling Hamiltonian. It is now possible to provide an overview for the generation of a tunable coupling between the JPA modes: the frequencies are designed so that the JPA modes are dispersively coupled only to the the $x$-modes, and a classical drive of frequency $\omega_d$ on the far detuned $z$-mode triggers the four-body coupling. The $z$-mode is driven by applying fields with equal strength but opposite phases to the capacitors $C_x$ and $C_y$ as indicated by the dark and light arrows in Supplementary Figure **6**. Since there is no drive to the $y$-mode, it can be dropped from the Hamiltonian. As a result, the total Hamiltonian can be written as

$$\hat{H} = \sum_{k=1}^{4} \left[ \hat{H}_{\text{JPA},k} + g_k^x(\hat{a}_l^\dagger \hat{a}_x + \hat{a}_x^\dagger \hat{a}_k) + g_k^z(\hat{a}_k^\dagger \hat{a}_z + \hat{a}_z^\dagger \hat{a}_k) \right] + E_J \frac{\hat{\phi}_x^2}{2\phi_0^3} \hat{\phi}_z - E_J \frac{\hat{\phi}_x^4}{96\phi_0^5} \hat{\phi}_z. \tag{S26}$$

Furthermore, if $\Delta_k = \omega_x - \omega_{r,k} \gg g_k$, then it is possible to apply a dispersive unitary transformation $\hat{U} = \exp(-i(g_1/\Delta_1)\hat{a}_x^\dagger \hat{a}_1 - i(g_2/\Delta_2)\hat{a}_x^\dagger \hat{a}_2 + i(g_3/\Delta_3)\hat{a}_x^\dagger \hat{a}_3 + i(g_4/\Delta_4)\hat{a}_x^\dagger \hat{a}_4 + \text{h.c.})$ to the total Hamiltonian. As mentioned before, the $z$-mode is driven classically by a field of frequency $\omega_d$ and we therefore replace $\hat{\phi}_z$ by the classical amplitude $2\phi_z\sqrt{n}\cos(\omega_d t)$, where $n$ is the number of photons in the mode $z$. If $\omega_d = \omega_{p,1} + \omega_{p,2} + \omega_{p,3} - \omega_{p,4}$ then it is possible to eliminate the fast rotating terms and the resulting Hamiltonian is

$$\hat{H}_{\text{plaquette}} \approx \sum_{k=1}^{4} \left( \hat{H}_{\text{JPA},k} - \frac{(g_k^x)^2}{\Delta_k^x} \hat{a}_k^\dagger \hat{a}_k \right) - C_{\text{jrm}}(\hat{a}_1^\dagger \hat{a}_2^\dagger \hat{a}_3^\dagger \hat{a}_4 + \text{h.c.}), \tag{S27}$$

where

$$C_{\text{jrm}} = E_J \sqrt{n} \frac{\phi_x^4 \phi_z}{4\phi_0^5} \frac{g_1 g_2 g_3 g_4}{\Delta_1 \Delta_2 \Delta_3 \Delta_4}. \tag{S28}$$

Note that in the computational basis the interaction term $C_{\text{jrm}}(\hat{a}_1^\dagger \hat{a}_2^\dagger \hat{a}_3^\dagger \hat{a}_4 + \text{h.c.})$ can be expressed as $2C_{\text{jrm}}\text{Re}[\alpha_0^3 \alpha_0^*]\hat{\sigma}_{z,1}\hat{\sigma}_{z,2}\hat{\sigma}_{z,1}\hat{\sigma}_{z,3}\hat{\sigma}_{z,4}$, which is the desired four-body coupling. By choosing for $E_J/(2\pi) = 860$ GHz, $g_k/\Delta_k = 0.12$, $\phi_x/\phi_0 = \phi_z/\phi_0 = 0.12$ and $n = 2.25$, it is possible to obtain a four-body coupling strength $C_{\text{jrm}} = 2\pi \times 1.7$ KHz. Note that this is a higher order coupling compared with the fixed one analyzed in Supplementary Note 4 and therefore is an order of magnitude smaller. The advantage of this scheme, however, is that it is tunable and the coupling strength can be increased by increasing $\sqrt{n}$, which is proportional to the strength of the applied microwave drive.

---

 

\end{document}